\documentclass[12pt]{article}
\usepackage{amsmath,amssymb}
\usepackage{bm}
\usepackage{color}
\usepackage{cite}
\usepackage{mathtools}

\usepackage{enumerate}
\usepackage{braket}

\usepackage{here}
\usepackage{comment}

\def\slash#1{\not\!\!#1}

\begin{document}

\title{
\begin{flushright}
\ \\*[-80pt]
\begin{minipage}{0.2\linewidth}
\normalsize
EPHOU-23-016\\*[50pt]
\end{minipage}
\end{flushright}
{\Large \bf
Modular symmetry in magnetized $T^{2g}$ torus \\ 
and orbifold models
\\*[20pt]}}

\author{
~Shota Kikuchi$^1$,
~Tatsuo Kobayashi$^1$,
~Kaito Nasu$^1$, \\
~Shohei Takada$^1$, and
~Hikaru Uchida$^2$
\\*[20pt]
\centerline{
\begin{minipage}{\linewidth}
\begin{center}
{\it \normalsize
$^1$Department of Physics, Hokkaido University, Sapporo 060-0810, Japan \\
$^2$Institute for the Advancement of Graduate Education, Hokkaido Unicersity, Sapporo 060-0817, Japan} \\*[5pt]
\end{center}
\end{minipage}}
\\*[50pt]}

\date{
\centerline{\small \bf Abstract}
\begin{minipage}{0.9\linewidth}
\medskip
\medskip
\small
We study the modular symmetry in magnetized $T^{2g}$ torus  and orbifold models.
The $T^{2g}$ torus has the modular symmetry $\Gamma_{g}=Sp(2g,\mathbb{Z})$.
Magnetic flux background breaks the modular symmetry 
to a certain normalizer $N_{g}(H)$.
We classify remaining modular symmetries by magnetic flux matrix types.
Furthermore, we study the modular symmetry for wave functions on the magnetized $T^{2g}$ and certain orbifolds.
It is found that wave functions on magnetized $T^{2g}$ as well as its orbifolds behave as the Siegel modular forms of weight $1/2$ and $\widetilde{N}_{g}(H,h)$, which is the metapletic congruence subgroup of the double covering group of $N_{g}(H)$, $\widetilde{N}_{g}(H)$.
Then, wave functions transform non-trivially under the quotient group, $\widetilde{N}_{g,h}=\widetilde{N}_{g}(H)/\widetilde{N}_{g}(H,h)$, where the level $h$ is related to the determinant of the magnetic flux matrix.
Accordingly, the corresponding four-dimensional (4D) chiral fields also transform non-trivially under $\widetilde{N}_{g,h}$ modular flavor transformation with modular weight $-1/2$.
We also study concrete modular flavor symmetries of wave functions on magnetized $T^{2g}$ orbifolds.
\end{minipage}
}

\begin{titlepage}
\maketitle
\thispagestyle{empty}
\end{titlepage}

\newpage


\section{Introduction}
\label{Intro}

One of the significant mysteries in the particle physics is the origin of flavor structure of  quarks and leptons such as their hierarchical masses, flavor mixing angles and CP phases.
Many scenarios have been studied.
Among them, one of the interesting approaches is to assume  certain non-Abelian discrete flavor symmetries such as $S_N$, $A_N$, $\Delta(3N^2)$, and $\Delta(6M^2)$ among  generations of quarks and leptons~\cite{
Altarelli:2010gt,Ishimori:2010au,Kobayashi:2022moq,Hernandez:2012ra,King:2013eh,King:2014nza}.

As the origin of flavor symmetries, the geometrical symmetries of compact spaces predicted in higher dimensional theories such as superstring theory have been attractive. 
(See e.g. Refs.~\cite{Kobayashi:2006wq,Abe:2009vi}.) 
The modular symmetry is the geometrical symmetry of compact spaces 
like tori, orbifolds and Calabi-Yau manifolds as transformation of 
cycle basis.
Recently, the modular  symmetry has been attractive since it includes 
certain non-Abelian discrete flavor symmetries like $S_3$, $A_4$, $S_4$, $A_5$ 
\cite{deAdelhartToorop:2011re}.
Thus, the modular symmetry can be a source of flavor symmetries 
of quarks and leptons, obtained from the compactification.
Actually, various bottom-up approaches of model building have been studied~\cite{
Feruglio:2017spp,Kobayashi:2018vbk,Penedo:2018nmg,Criado:2018thu,Kobayashi:2018scp,Novichkov:2018ovf,Novichkov:2018nkm,deAnda:2018ecu,Okada:2018yrn,Kobayashi:2018wkl,Novichkov:2018yse}, 
in which the assumed modular flavor symmetries play an important role to determine the flavor structure of quarks and leptons from the geometrical parameters called moduli.
(See for more references Ref.~\cite{Kobayashi:2023zzc}.)

In top-down approaches, on the other hand, it is important to look for what modular flavor symmetries of  chiral fields such as quarks and leptons appear in an effective theory of superstring theory.
For example, ten-dimensional (10D) ${\cal N}=1$ supersymmetric Yang-Mills theory with non-vanishing magnetic fluxes on a torus or its orbifold is an interesting effective theory of magnetized D-brane models in superstring theory~\cite{
Bachas:1995ik,Blumenhagen:2000wh,Angelantonj:2000hi,Blumenhagen:2000ea}.
Magnetized D-brane models have several interesting features.
Multi-generational chiral fermions~\cite{
Cremades:2004wa,Abe:2008fi,Abe:2013bca,Abe:2014noa,Kobayashi:2017dyu,Sakamoto:2020pev,Antoniadis:2009bg,Kikuchi:2022lfv,Kikuchi:2022psj,Kikuchi:2023awm}
(including three generations of chiral fermions~\cite{
Abe:2008sx,Abe:2015yva,Hoshiya:2020hki})
can be obtained by specific magnetic fluxes and boundary conditions.
Their Yukawa couplings~\cite{Cremades:2004wa,Fujimoto:2016zjs,Antoniadis:2009bg,Kikuchi:2022lfv}
as well as higher-order couplings~\cite{Abe:2009dr} and also Majorana neutrino mass terms generated by D-brane instanton effects~\cite{Kobayashi:2015siy,Hoshiya:2021nux} can be calculated analytically since we can find their wave functions explicitly.
(See for D-brane instanton computations Refs.~\cite{Blumenhagen:2006xt, Ibanez:2006da}.)
Actually, realistic quark and lepton masses and mixing angles as well as CP phases have been realized in Refs.~\cite{
Abe:2012fj,Abe:2014vza,Fujimoto:2016zjs,Kobayashi:2016qag,Buchmuller:2017vho,Buchmuller:2017vut,Kikuchi:2021yog,Kikuchi:2022geu,Hoshiya:2022qvr}.

These magnetized torus and orbifold models have the modular symmetry.
\footnote{For heterotic orbifold models, see Refs.~\cite{Lauer:1989ax,Lerche:1989cs,Ferrara:1989qb,Baur:2019kwi,Nilles:2020nnc}.}
Their effective field theory is controlled by the modular symmetry.
The torus $T^2$ has a single complex structure modulus and the modular symmetry $SL(2,\mathbb{Z})$.
The modular symmetry and its implications on the flavor structure in 
the magnetized $T^2$ and its certain orbifold models with a single complex structure modulus 
have been studied in Refs.~\cite{
Kobayashi:2017dyu,Kobayashi:2018rad}
and in detail in Refs.~\cite{Kikuchi:2020frp,Kikuchi:2021ogn}. (See also Refs.~\cite{
Kobayashi:2016ovu,Kobayashi:2018bff,Kariyazono:2019ehj,Kobayashi:2020hoc,Ohki:2020bpo,Kikuchi:2020nxn,Almumin:2021fbk,Tatsuta:2021deu,Kikuchi:2022bkn}.)

Generic compact space has many moduli and they have a larger modular symmetry, 
$Sp(2g,\mathbb{Z})$.
That can lead to a rich flavor structure.
Such symplectic modular symmetries were studied in Calabi-Yau compactifications  
\cite{Strominger:1990pd,Candelas:1990pi, Ishiguro:2020nuf, Ishiguro:2021ccl}
and others 
\cite{Baur:2020yjl, Nilles:2021glx, Ding:2020zxw, Ding:2021iqp}.
The torus $T^{2g}$ has the modular symplectic symmetry $Sp(2g,\mathbb{Z})$.
In this paper, we extend the above analysis on magnetized $T^2$ and orbifold models 
to the magnetized $T^{2g}$ (including $T^4$ and $T^6$) and its certain orbifold models.
We study the modular flavor symmetry in these higher dimensional torus and 
orbifold models with magnetic fluxes.


This paper is organized as follows.
We review the modular symmetry on $T^{2g}$ in subsection~\ref{subsec:T2gMS} and modular forms in subsection~\ref{subsec:MF}.
In section~\ref{sec:MT2g}, we review magnetized $T^{2g}$ compactification.
Then, we study the modular symmetry in magnetized $T^{2g}$ and orbifold compactifications.
In particular, we classify the modular symmetry, which is consistent with a magnetic flux in subsection~\ref{subsec:MSMt2g}, and then we study the modular flavor symmetry of wave functions on magnetized $T^{2g}$ as well as orbifold by general analysis and concrete examples in each class, in subsection~\ref{subsec:MSWMT2gOM}.
In section,~\ref{conclusion}, we conclude this study.
In Appendix~\ref{ap:STn}, we discuss the algebraic relation between $S$ transformation and general $T$ transformation.
In Appendix~\ref{ap:Landsberg-Schaar}, we derive the Landsberg-Schaar relation.
In Appendix~\ref{ap:proof}, we prove that the generators of $\Delta(96) \times Z_4$ satisfy the algebraic relation.
In Appendix~\ref{ap:OmegatauI}, we discuss the modular flavor symmetry with the moduli which are proportional to the unit matrix.


\section{Modular symmetry on $T^{2g}$ and modular forms}
\label{sec:MST2gMF}

In this section, we review the modular symmetry on $T^{2g}$ and modular forms. (See e.g. Refs.~\cite{Siegel:1943,Maass:1971,Igusa:1972,Freitag:1983,Freitag:1991,Klingen:1990,Geer:2008,Fay:1973,Mumford:1984,Ding:2020zxw}.)
A $2g$-dimensional torus $T^{2g}$ can be constructed as $\mathbb{C}^{g}/\Lambda$, where $\Lambda$ is a lattice on $\mathbb{C}^{g}$ spanned by $2g$ numbers of lattice vectors $\alpha_j$ and $\beta_j$ ($j=1,...,g$).
We write complex coordinates of $\alpha_j$ and $\beta_j$ as $\alpha_j = {^t}(\alpha_{1j},...,\alpha_{gj})$ and $\beta_j = (\beta_{1j},...,\beta_{gj})$, respectively.
Then, the $a$th component of the complex coordinate on $\mathbb{C}^{g}$, $u^a$, can be written as
\begin{align}
u^a = \sum_{j=1}^{g} \alpha^a_j x^j + \beta^a_j y^j \quad (x^j, y^j \in \mathbb{R}).
\end{align}
We also define the $j$th component of the complex coordinate on $T^{2g}$, $z^j$, as
\begin{align}
z^j = (\alpha^{-1})^j_a u^a = x^j + \Omega^j_k y^k \quad (0 \leq x^j, y^k \leq 1).
\end{align}
Here, we consider that the $g \times g$ complex matrix, $\Omega = \alpha^{-1}\beta$, is symmetric and the eigenvalues of the imaginary part is positive,~i.e.,
\begin{align}
{^t}\Omega = \Omega, \quad {\rm Im}\Omega > 0, \label{Omega}
\end{align}
which is so-called the complex structure moduli.
Then, the lattice identification is written by
\begin{align}
z+e_k \sim z+\Omega e_k \sim z,
\end{align}
for $\forall k$, where $j$th component of $e_k$ is $\delta_{j,k}$.
The metric of $T^{2g}$ is given by
\begin{align}
ds^2 = d\bar{u}^a du^a = (\alpha^{\dagger}\alpha)_{ij} d\bar{z}^i dz^j = 2h_{ij} d\bar{z}^i dz^j,
\end{align}
and then the volume of $T^{2g}$ can be calculated as
\begin{align}
{\rm Vol}(T^{2g}) &= \int_{T^{2g}} d^gzd^g\bar{z} \sqrt{|{\rm det}(2h)|} = |{\rm det}(\alpha^{\dagger}\alpha)|^2 {\rm det}(2{\rm Im}\Omega).
\end{align}
The gamma matrices on $T^{2g}$, $\Gamma^{\bar{z}^i}$ and $\Gamma^{z^j}$, satisfying $\{ \Gamma^{\bar{z}^i}, \Gamma^{z^j} \} = 2h^{ij}$, are given by
\begin{align}
\Gamma^{\bar{z}^i} = [(\alpha^{\dagger})^{-1}]^i_a \Gamma^{\bar{u}^a}, \quad \Gamma^{z^j} = [\alpha^{-1}]^j_b \Gamma^{u^b},
\end{align}
where $\Gamma^{\bar{u}^a}$ and $\Gamma^{u^b}$ are the gamma matrices on $\mathbb{C}^{g}$, satisfying $\{ \Gamma^{\bar{u}^a}, \Gamma^{u^b} \} = 2\delta^{a,b}$.


\subsection{Modular symmetry on $T^{2g}$}
\label{subsec:T2gMS}

Now, let us review the Siegel modular symmetry on $T^{2g}$.
Let us consider the following lattice transformation,
\begin{align}
\begin{pmatrix}
{^t}\gamma(\beta) \\ {^t}\gamma(\alpha)
\end{pmatrix}
=
\begin{pmatrix}
A & B \\
C & D
\end{pmatrix}
\begin{pmatrix}
{^t}\beta \\ {^t}\alpha
\end{pmatrix}
=
\begin{pmatrix}
A{^t}\beta+B{^t}\alpha \\ C{^t}\beta+D{^t}\alpha
\end{pmatrix}, \quad
\gamma =
\begin{pmatrix}
A & B \\
C & D
\end{pmatrix} \in Sp(2g,\mathbb{Z}) \equiv \Gamma_{g},
\label{latticetransf}
\end{align}
where $g \times g$ matrices, $A$, $B$, $C$, $D$, satisfy
\begin{align}
{^t}AC={^t}CA, \ {^t}BD={^t}DB, \ {^t}AD-{^t}CB=I_{g}.
\end{align}
Since the lattice spanned by $\gamma(\alpha)$ and $\gamma(\beta)$ is the same as the lattice spanned by $\alpha$ and $\beta$, there is $\Gamma_{g}=Sp(2g,\mathbb{Z})$ symmetry in $T^{2g}$ compactification.
Associated with the lattice transformations in Eq.~(\ref{latticetransf}), the moduli $\Omega$  and the coordinate $z$ transform as
\begin{align}
&\gamma: \Omega = \alpha^{-1}\beta \rightarrow \gamma(\Omega) = \gamma(\alpha)^{-1}\gamma(\beta) = (A\Omega+B)(C\Omega+D)^{-1}, \label{gammaOmega} \\
&\gamma: z = \alpha^{-1}u \rightarrow \gamma(z) = \gamma(\alpha)^{-1}u = {^t}(C\Omega+D)^{-1}z.
\label{gammaz}
\end{align}
Eq.~(\ref{gammaOmega}) is called the (inhomogeneous) Siegel modular transformation.
Here, we call Eqs.~(\ref{gammaOmega}) and (\ref{gammaz}) the Siegel modular transformation.

The generators of the Siegel modular group, $\Gamma_{g}=Sp(2g,\mathbb{Z})$, are
given by 
\begin{align}
S =
\begin{pmatrix}
0 & I_{g} \\
-I_{g} & 0
\end{pmatrix}, \quad
T_{ab} =
\begin{pmatrix}
I_{g} & B_{ab} \\
0 & I_{g}
\end{pmatrix},
\label{Sp(2g,Z)gen}
\end{align}
where $B_{ab}$ are concretely written as
\begin{align}
B_{11} = 1,
\end{align}
for $g=1$,
\begin{align}
B_{11} =
\begin{pmatrix}
1 & 0 \\
0 & 0
\end{pmatrix}, \ 
B_{22} =
\begin{pmatrix}
0 & 0 \\
0 & 1
\end{pmatrix}, \ 
B_{12} =
\begin{pmatrix}
0 & 1 \\
1 & 0
\end{pmatrix},
\end{align}
for $g=2$, and
\begin{align}
\begin{array}{c}
B_{11} =
\begin{pmatrix}
1 & 0 & 0\\
0 & 0 & 0 \\
0 & 0 & 0
\end{pmatrix}, \ 
B_{22} =
\begin{pmatrix}
0 & 0 & 0 \\
0 & 1 & 0 \\
0 & 0 & 0
\end{pmatrix}, \ 
B_{33} =
\begin{pmatrix}
0 & 0 & 0 \\
0 & 0 & 0 \\
0 & 0 & 1
\end{pmatrix}, \\
B_{12} =
\begin{pmatrix}
0 & 1 & 0 \\
1 & 0 & 0 \\
0 & 0 & 0
\end{pmatrix}, \ 
B_{13} =
\begin{pmatrix}
0 & 0 & 1 \\
0 & 0 & 0 \\
1 & 0 & 0
\end{pmatrix}, \ 
B_{23} =
\begin{pmatrix}
0 & 0 & 0 \\
0 & 0 & 1 \\
0 & 1 & 0
\end{pmatrix},
\end{array}
\end{align}
for $g=3$.
The generators in Eq.~(\ref{Sp(2g,Z)gen}) satisfy the following relations,
\begin{align}
\begin{array}{l}
S^2 =
\begin{pmatrix}
-I_{g} & 0 \\
0 & -I_{g}
\end{pmatrix}
=-I_{2g} \equiv R, \\
S^4 = R^2 = I_{2g}, \\
(ST_{ab})^3 =
\begin{pmatrix}
B_{ab} & -I_{g} + B_{ab}^2 \\
I_{g} - B_{ab}^2 & B_{ab}
\end{pmatrix}, \\
(ST_{ab})^6 =
\begin{pmatrix}
-I_{g} + 2B_{ab}^2 & 0 \\
0 & -I_{g} + 2B_{ab}^2
\end{pmatrix}
= A_{-I_{g}+2B_{ab}^2} \equiv U, \\
(ST_{ab})^{12} = I_{2g}, \\
RX=XR \ (X=S, T_{ab}),
\end{array}
\label{algSP(2g,Z)}
\end{align}
where we denote
\begin{align}
A_{X} \equiv
\begin{pmatrix}
X & 0 \\
0 & {^t}X^{-1}
\end{pmatrix} \in Sp(2g,\mathbb{Z}), \quad X \in GL(g,\mathbb{Z}),
\end{align}
and then $A_{\pm I_{g}}=\pm I_{2g}$.
Note that Eq.~(\ref{algSP(2g,Z)}) corresponds to $S^3$ transformation in the space spanned by lattice vectors $\alpha_c$ and $\beta_c$ ($c \neq a, b$).
In Appendix~\ref{ap:STn}, we discuss the algebraic relation between $S$ transformation and general $T$ transformation generated by $T_{ab}$ transformation.
Under the $S$ and $T_{ab}$ transformations, $z$ and $\Omega$ transform as
\begin{align}
S: (z, \Omega) \rightarrow (\gamma(z, \Omega)) = (-\Omega^{-1}z, -\Omega^{-1}), \quad
T_{ab} :(z, \Omega) \rightarrow (\gamma(z, \Omega)) = (z, \Omega+B_{ab}).
\label{STabzOmega}
\end{align}


\subsection{Modular forms}
\label{subsec:MF}

Now, let us review the Siegel modular forms.
First, we introduce the principal congruence subgroup of level $n$,
\begin{align}
\Gamma_{g}(n) = \left\{ \gamma' =
\begin{pmatrix}
A' & B' \\
C' & D'
\end{pmatrix} \biggl|
\begin{pmatrix}
A' & B' \\
C' & D'
\end{pmatrix}
\equiv
\begin{pmatrix}
I_{g} & 0 \\
0 & I_{g}
\end{pmatrix} \quad
({\rm mod}\ n) \right\},
\end{align}
in particular, $\Gamma_{g}(1)=\Gamma_{g}$.
The Siegel modular forms $f(\Omega)$ of integral weight $k$ and level $n$ at genus $g$ are holomorphic functions of $\Omega$ which satisfy
\begin{align}
&\gamma:f(\Omega) \rightarrow f(\gamma(\Omega))=J_k(\gamma,\Omega)\rho(\gamma)f(\Omega), \ 
J_k(\gamma,\Omega)=[{\rm det}(C\Omega+D)]^k, \ 
\gamma=
\begin{pmatrix}
A & B \\
C & D
\end{pmatrix}
\in \Gamma_{g}, \\
&J_k(\gamma_2\gamma_1,\Omega) = J_k(\gamma_2, \gamma_1(\Omega)) J_k(\gamma_1, \omega), \quad
\rho(\gamma_2\gamma_1) = \rho(\gamma_2)\rho(\gamma_1), \quad \gamma_1, \gamma_2 \in \Gamma_{g},
\end{align}
with
\begin{align}
\rho(\gamma')=I, \quad \gamma' \in \Gamma_{g}(n).
\end{align}
Thus, $\rho$ is a unitary representation of the quotient group, $\Gamma_{g,n} = \Gamma_{g}/\Gamma_{g}(n)$, so-called the finite Siegel modular group.
In other words, the Siegel modular forms transform non-trivially under the finite Siegel modular transformation, $\Gamma_{g,n}$.
Concretely, we obtain the relations,
\begin{align}
\begin{array}{c}
\rho(R)^2=\rho(S)^4=[ \rho(S)\rho(T_{ab}) ]^{12}=\rho(T_{ab})^n=I, \\ 
(\rho(R)=\rho(S)^2=[ \rho(S)\rho(T_{ab}) ]^{6}=I \ (n=2)),
\end{array}
\end{align}
and also
\begin{align}
\rho(R)=\rho(S)^2=(-1)^{gk}I \quad \left( \Leftarrow f(\Omega)=f(S^2(\Omega))=(-1)^{gk} \rho(S)^2 f(\Omega) \right).
\end{align}
Then, we also find that
\begin{align}
\rho(R)\rho(X)=\rho(X)\rho(R) \quad (X=S, T_{ab}).
\end{align}
On the other hand, the overall factor $J_k(\gamma,\Omega)$ is called the automorphy factor, and it can be uniquely determined once $\gamma$ is given.

Here, we comment on the stabilizer, $H$, and the normalizer, $N_{g}(H)$, mentioned in Ref.~\cite{Ding:2020zxw}.
When the moduli $\Omega$ are restricted to a certain region, as shown in section~\ref{subsec:MSMt2g}, the modular symmetry is reduced from the Siegel modular group, $\Gamma_{g}$, to the normalizer $N_{g}(H)$, called the Siegel modular subgroup.
When $\Omega$ is fixed to a certain form in the region, 
the stabilizer $H$ is the unbroken group which is generated by the modular transformation such that $\Omega$ is invariant.
In general, the stabilizer is a normal subgroup of $N(H)$.
As with $\Gamma_{g}(n)$ and $\Gamma_{g,n}=\Gamma_{g}/\Gamma_{g}(n)$, we can consider the principal congruence subgroup of $N(H)$,
\begin{align}
N_{g}(H,n) \equiv \{ \gamma' \in N_{g}(H) | \gamma' \equiv I_{2g} \ ({\rm mod}\ n) \},
\end{align} 
and the quotient group, $N_{g,n}(H) = N_{g}(H)/N_{g}(H,n)$, called the finite Siegel modular subgroup.

We extend the above analysis for wave functions $\psi(z,\Omega)$ in section~\ref{subsec:MSWMT2gOM}.
To see that, however, we have to introduce the Siegel modular forms of half-integral weight.
First, we introduce the metapletic double covering group of $\Gamma_{g} = Sp(2g,\mathbb{Z})$,
\begin{align}
\widetilde{\Gamma}_{g} = \widetilde{Sp}(2g,\mathbb{Z}) = \left\{ \widetilde{\gamma} = [\gamma, \epsilon] | \gamma \in \Gamma_{g}, \epsilon \in \{\pm1\} \right\}. \label{tildeGammag}
\end{align}
The multiplication is given by
\begin{align}
\widetilde{\gamma}_1 \widetilde{\gamma}_2 = [\gamma_1, \epsilon_1] [\gamma_2, \epsilon_2] = [\gamma_1\gamma_2, A(\gamma_1,\gamma_2) \epsilon_1 \epsilon_2] \equiv [\gamma_{1,2}, \epsilon_{1,2}] = \widetilde{\gamma}_{1,2} \in \widetilde{\Gamma}_{g}, \label{eq:multipl}
\end{align}
where $A(\gamma_1,\gamma_2)$ denotes the Rao's 2-cocycle~\cite{Rao:1993}\footnote{See also Ref.~\cite{Szpruch:2007}.}, satisfying the following relation,
\begin{align}
(\widetilde{\gamma}_1\widetilde{\gamma}_2)\widetilde{\gamma}_3 &= \widetilde{\gamma}_1 (\widetilde{\gamma}_2\widetilde{\gamma}_3), \notag \\
\Leftrightarrow \ 
A(\gamma_1,\gamma_2) A(\gamma_1\gamma_2, \gamma_3) &= A(\gamma_1, \gamma_2\gamma_3) A(\gamma_2,\gamma_3). \label{cocyclerelation}
\end{align}
In particular, the generators of $\widetilde{\Gamma}_{g} = \widetilde{Sp}(2,\mathbb{Z})$ are given by 
\begin{align}
\widetilde{S} \equiv [S,(-1)^{g}], \quad \widetilde{T}_{ab} \equiv [T_{ab},1], \label{StildeTtilde}
\end{align}
and they satisfy the following relations;
\begin{align}
\begin{array}{l}
\widetilde{S}^2 = [-I_{2g},1] \equiv \widetilde{R}, \quad \widetilde{R}^2 = \widetilde{S}^4 = [I_{2g},(-1)^{g}], \quad \widetilde{R}^4 = \widetilde{S}^8 = [I_{2g},1] \equiv \widetilde{I}_{2g}, \\
(\widetilde{S}\widetilde{T}_{ab})^{3} = [(ST_{ab})^3,(-1)^{g'-1}], \quad (\widetilde{S}\widetilde{T}_{ab})^{6} = [A_{-I_{g}+2B_{ab}^2}, (-1)^{g-1}], \\
(\widetilde{S}\widetilde{T}_{ab})^{12} = [I_{2g},(-1)^{g-g'}] \equiv \widetilde{U}, \quad \widetilde{U}^2 = (\widetilde{S}\widetilde{T}_{ab})^{24} = [I_{2g},1] = \widetilde{I}_{2g}, \\
\widetilde{R}\widetilde{X} = \widetilde{X}\widetilde{R}, \quad \widetilde{U}\widetilde{X} = \widetilde{X}\widetilde{U}, \quad (\widetilde{X}=\widetilde{S}, \widetilde{T}_{ab}),
\end{array}
\label{algtildeSL2Z}
\end{align}
with
\begin{align}
g'= \left\{
\begin{array}{ll}
1 & (a=b) \\
2 & (a \neq b)
\end{array}
\right..
\label{g'}
\end{align}
Similarly, we can introduce the metapletic congruence subgroup of level $n(\in 4\mathbb{Z})$,
\begin{align}
\widetilde{\Gamma}_{g}(n) = \{ \widetilde{\gamma}'=[\gamma',\epsilon] \in \widetilde{\Gamma}_{g} | \gamma' \in \Gamma_{g}(n), \epsilon=1 \}. \label{tildeGammagn}
\end{align}
Then, the Siegel modular forms $f(\Omega)$ of half-integral weight $k/2$ and level $n$ at genus $g$ are holomorphic functions of $\Omega$ which satisfy
\begin{align}
&\widetilde{\Gamma}_{g} \ni \widetilde{\gamma}:f(\Omega) \rightarrow f(\widetilde{\gamma}(\Omega))=\widetilde{J}_{k/2}(\widetilde{\gamma},\Omega)\rho(\widetilde{\gamma})f(\Omega), \ 
\widetilde{J}_{k/2}(\widetilde{\gamma},\Omega)=\epsilon^kJ_{k/2}(\gamma,\Omega)=\epsilon^k[{\rm det}(C\Omega+D)]^{k/2}, \\
&\widetilde{J}_{k/2}(\widetilde{\gamma}_2\widetilde{\gamma}_1,\Omega) = [A(\gamma_1,\gamma_2)]^k \widetilde{J}_{k/2}(\widetilde{\gamma}_2, \widetilde{\gamma}_1(\Omega)) \widetilde{J}_{k/2}(\widetilde{\gamma}_1, \omega), \quad 
\rho(\widetilde{\gamma}_2\widetilde{\gamma}_1) = \rho(\widetilde{\gamma}_2)\rho(\widetilde{\gamma}_1), \quad \widetilde{\gamma}_1, \widetilde{\gamma}_2 \in \widetilde{\Gamma}_{g},
\end{align}
with
\begin{align}
\rho(\widetilde{\gamma}')=I, \quad \widetilde{\gamma}' \in \widetilde{\Gamma}_{g}(n).
\end{align}
Here, we note that $\widetilde{\gamma}(\Omega)=\gamma(\Omega)$ and we choose $(-1)^{gk/2}=e^{-\pi igk/2}$.
Thus, $\rho$ is a unitary representation of the quotient group, $\widetilde{\Gamma}_{g,n} = \widetilde{\Gamma}_{g}/\widetilde{\Gamma}_{g}(n)$, called the metapletic finite Siegel modular group.
In other words, the Siegel modular forms transform non-trivially under the metapletic finite Siegel modular transformation, $\widetilde{\Gamma}_{g,n}$.
Concretely, we obtain the relations,
\begin{align}
\begin{array}{c}
\rho(\widetilde{R})^4=\rho(\widetilde{S})^8=I, \ \rho(\widetilde{U})^2=[ \rho(\widetilde{S})\rho(\widetilde{T}_{ab}) ]^{24}=I, \ \rho(\widetilde{T}_{ab})^n=I, \\ 
(\rho(\widetilde{R})=\rho(\widetilde{S})^2=I \ (n=2)),
\end{array}
\label{rhoI}
\end{align}
and also
\begin{align}
&\rho(\widetilde{R})=\rho(\widetilde{S})^2=e^{\pi igk/2}I \quad \left( \Leftarrow f(\Omega)=f(\widetilde{S}^2(\Omega))=e^{-\pi igk/2} \rho(\widetilde{S})^2 f(\Omega) \right), \label{rhoRtilde} \\
&\rho(\widetilde{R})^2=\rho(\widetilde{S})^4=(-1)^{gk}I \quad \left( \Leftarrow f(\Omega)=f(\widetilde{S}^4(\Omega))=(-1)^{gk} \rho(\widetilde{S})^4 f(\Omega) \right), \label{rhoR2tilde} \\
&\rho(\widetilde{U})=\rho(\widetilde{S}\widetilde{T}_{ab})^{12}=(-1)^{(g-g')k}I \quad \left( \Leftarrow f(\Omega)=f((\widetilde{S}\widetilde{T}_{ab})^{12}(\Omega))=(-1)^{(g-g')k} \rho(\widetilde{S}\widetilde{T}_{ab})^{12} f(\Omega) \right), \label{rhoUtilde}
\end{align}
Then, we also find that
\begin{align}
\rho(\widetilde{R})\rho(\widetilde{X})=\rho(\widetilde{X})\rho(\widetilde{R}), \quad \rho(\widetilde{U})\rho(\widetilde{X})=\rho(\widetilde{X})\rho(\widetilde{U}), \quad (\widetilde{X}=\widetilde{S}, \widetilde{T}_{ab}).
\label{RXXR}
\end{align}
On the other hand, the overall factor $J_k(\gamma,\Omega)$, called the automorphy factor, can be uniquely determined once $\widetilde{\gamma}$ is given.

Similarly, we can consider the double covering group of $N_{g}(H)$, $\widetilde{N}_{g}(H)$, and the metapletic congruence subgroup of level $n(\in 4\mathbb{Z})$, $\widetilde{N}_{g}(H,n)$, by replacing $\gamma \in \Gamma_{g}$ in Eq.~(\ref{tildeGammag}) with $\gamma \in N_{g}(H)$ and replacing $\gamma' \in \Gamma_{g}(n)$ in Eq.~(\ref{tildeGammagn}) with $\gamma' \in N_{g}(H,n)$, respectively.
In that case, $\rho$ is a unitary representation of the quotient group, $\widetilde{N}_{g,n}(H) = \widetilde{N}_{g}(H)/\widetilde{N}_{g}(H,n)$, called the metapletic finite Siegel modular subgroup.


\section{Magnetized $T^{2g}$ compactification}
\label{sec:MT2g}

In this section, we review on a $T^{2g}$ compactification with a background magnetic flux, so-called a magnetized $T^{2g}$ compactification~\cite{Cremades:2004wa,Antoniadis:2009bg}.
We introduce the following $U(1)$ magnetic flux,
\begin{align}
F = \pi [ {^t}N ({\rm Im}\Omega)^{-1} ]_{ij} (i dz^i \land d\bar{z}^j), \label{F}
\end{align}
with
\begin{align}
{^t}(N\Omega)=N\Omega, \label{Fflat}
\end{align}
which is needed for the magnetic flux $F$ to be $(1,1)$ form (F-flat condition).
Here, $N$ denotes $g \times g$ flux matrix and the flux must be quantized,~i.e. $N_{ij} \in \mathbb{Z}$.
It is induced by the gauge potential,
\begin{align}
A(z) 
&= \pi {\rm Im}[ {^t}(N\bar{z}) ({\rm Im}\Omega)^{-1}dz ] \notag \\
&= - \frac{\pi i}{2}[ {^t}(N\bar{z}) ({\rm Im}\Omega)^{-1} ]_k dz^k + \frac{\pi i}{2} [ {^t}(Nz) ({\rm Im}\Omega)^{-1} ]_k d\bar{z}^k \\
&= A_{z^k} dz^k + A_{\bar{z}^k} d\bar{z}^k. \notag
\end{align}
Here, we do not consider Wilson lines.
Under the lattice translations, it transforms as
\begin{align}
A(z+e_k) &= A(z) + d[\pi {^t}N ({\rm Im}\Omega)^{-1} {\rm Im}z]_k = A(z) + d\chi_{e_k}(z), \\
A(z+\Omega e_k) &= A(z) + d[\pi {\rm Im}\{(N\bar{\Omega}) ({\rm Im}\Omega)^{-1} z \}]_k = A(z) + d\chi_{\Omega e_k}(z).
\end{align}
It corresponds to $U(1)$ gauge transformation.
Through the covariant derivative with  $U(1)$ unit charge, $q=1$,
\begin{align}
D
&= d - i A(z) \notag \\
&= \left( \partial_{z^k} - \frac{\pi}{2}[ {^t}(N\bar{z}) ({\rm Im}\Omega)^{-1} ]_k \right) dz^k + \left( \partial_{\bar{z}^k} + \frac{\pi}{2} [ {^t}(Nz) ({\rm Im}\Omega)^{-1} ]_k \right) d\bar{z}^k \notag \\
&= D_{z^k} dz^k + D_{\bar{z}^k} d\bar{z}^k, \notag
\end{align}
wave functions on the magnetized $T^{2g}$ with $q=1$, satisfy the following boundary conditions,
\begin{align}
\Psi(z+e_k,\Omega) &= e^{2\pi i {^t}\alpha^S e_k } e^{i\chi_{e_k}(z)} \Psi(z), \label{BCPsiz1} \\
\Psi(z+\Omega e_k,\Omega) &= e^{2\pi i {^t}\beta^S e_k} e^{i\chi_{\Omega e_k}(z)} \Psi(z), \label{BCPsiztau}
\end{align}
where $\alpha^S = {^t}(\alpha^S_1,...,\alpha^S_g)$ and $\beta^S = {^t}(\beta^S_1,...,\beta^S_g)$ with $0 \leq \alpha^S_k, \beta^S_k \leq 1$ are called as the Scherk-Schwarz (SS) phases.
Note that the Wilson line phases can be converted into the SS phases through an appropriate gauge transformation.
We consider solutions of the zero-mode Dirac equation,
\begin{align}
i\slash{D}\Psi(z,\Omega) = i (\Gamma^{z^k}D_{z^k} + \Gamma^{\bar{z}^k}D_{\bar{z}^k} ) \Psi(z,\Omega) = 0,
\end{align}
with the boundary conditions in Eqs.~(\ref{BCPsiz1}) and (\ref{BCPsiztau}).
When all eigenvalues of $N$ are positive in addition to Eq.~(\ref{Omega}), only the component of $\Psi(z,\Omega)$ whose chirality on $\forall z^k\ (k=1,...,g)$ is positive has ${\rm det}N$ number of degenerated zero modes,
\begin{align}
&\psi_{T^{2g}}^{(J+\alpha^S,\beta^S),N}(z,\Omega) \notag \\
=& [{\rm Vol}(T^{2g})]^{-1/2} ({\rm det}N)^{1/4} e^{-2\pi i{^t}(J+\alpha^S)N^{-1}\beta^S} e^{\pi i {^t}(Nz) ({\rm Im}\Omega)^{-1} {\rm Im}z} \vartheta
\begin{bmatrix}
{^t}(J+\alpha^S)N^{-1}\\
-{^t}\beta
\end{bmatrix}
(Nz, N\Omega),
\label{wavT2g}
\end{align}
for $\forall J \in \Lambda_N$,
where $\vartheta$ denotes the Riemann-theta function defined by
\begin{align}
\vartheta
\begin{bmatrix}
{^t}a\\
{^t}b
\end{bmatrix}
(\nu, \Omega)
=
\sum_{l\in \mathbb{Z}^g}
e^{\pi i {^t}(l+a)\Omega(l+a)}
e^{2\pi i {^t}(l+a)(\nu+b)},
\quad a, b \in \mathbb{R}^g, \quad \nu \in \mathbb{C}^g,
\end{align}
and $\Lambda_N$ denotes the lattice cell spanned by
\begin{align}
{^t}Ne_k \ (k=1,...,g).
\end{align}
This means that
\begin{align}
\psi_{T^{2g}}^{(J+{^t}Ne_k+\alpha^S,\beta^S),N}(z,\Omega) = \psi_{T^{2g}}^{(J+\alpha^S,\beta^S),N}(z,\Omega).
\end{align}
The normalization condition is given by
\begin{align}
\int_{T^{2g}} d^gzd^g\bar{z} \sqrt{|{\rm det}(2h)|} \left(\psi^{(J+\alpha^S,\beta^S),N}_{T^{2g}}(z,\Omega)\right)^* \psi^{(K+\alpha^S,\beta^S),N}_{T^{2g}}(z,\Omega) 
= [{\rm det}(2{\rm Im}\Omega)]^{-1/2} \delta_{\bar{J},K}.
\label{Normalization}
\end{align}

Finally, we give a comment on four-dimensional (4D) low energy effective field theory.
We assume that 4D ${\cal N}=1$ supersymmetry remains, although 
our results on modular flavor symmetries in the following sections are independent of 
whether supersymmetry remains or not.
Higher dimensional fields $\Phi(x,z)$ are decomposed as follows,
\begin{align}
\Phi(x,z)=\sum_{n,I} \varphi^I_n(x) \psi^I_n(z),
\end{align}
i.e., the Kaluza-Klein decomposition.
Here, $I$ denotes the degeneracy index for a fixed mass level $n$.
The lowest modes with $n=0$ are relevant to 4D effective field theory, 
although they may be degenerate.
We naturally assume the canonical kinetic term of $\Phi(x,z)$.
Then, we integrate the extra dimension so as to obtain K\"ahler potential of 
4D low-energy effective field theory,
\begin{align}
K(\varphi, \bar{\varphi}) = Z_{\bar{J}K} \varphi^{J}_0(x)^{\dagger} \varphi^{K}_0(x), \quad Z_{\bar{J}K} = [{\rm det}(2{\rm Im}\Omega)]^{-1/2} \delta_{\bar{J},K}.
\end{align}

In the following section, we study the modular symmetry in the magnetized $T^{2g}$ and its orbifold compactifications.
Since the field $\Phi(x,z)$ is invariant under the modular symmetry, 
the modular transformation of the 4D fields $\varphi^I_0(x)$ is 
inverse to one of $\psi^I_0(z)$ \cite{Kikuchi:2022txy}.\footnote{
Note that the modular transformation corresponds to the change of 
basis $\psi^I_0(z)$.}
Hereafter, we consider the modular transformation for the zero-mode wave functions in Eq.~(\ref{wavT2g}).


\section{Modular symmetry in magnetized $T^{2g}$ and orbifold compactifications}
\label{sec:MSMT2gO}

In this section, we study the modular symmetry in magnetized $T^{2g}$ and orbifold compactifications.


\subsection{Modular symmetry consistent with magnetized fluxes}
\label{subsec:MSMt2g}

First, in this subsection, let us see what kind of the modular symmetry is consistent with a magnetic flux matrix $N$.
By using Eqs.~(\ref{gammaOmega}), (\ref{gammaz}) and (\ref{Fflat}), we can find that Eq.~(\ref{F}) is invariant and Eq.~(\ref{Fflat}) is consistent for the modular transformation by $\gamma=
\begin{pmatrix}
A & B \\
C & D
\end{pmatrix}$, when the following conditions
\begin{align}
(C\Omega+D)^{-1} {^t}N (C\Omega+D) &= N \ \Rightarrow \ {^t}CN = CN, \ {^t}D=D{^t}N, \\
{^t}[ N(A\Omega+B)(C\Omega+D)^{-1} ] &=  N(A\Omega+B)(C\Omega+D)^{-1} \ \Rightarrow \ AN =NA, \ B{^t}N = NB,
\end{align}
are satisfied.
In particular, in order for $N$ matrix to be consistent with generators in Eq.~(\ref{Sp(2g,Z)gen}), the following relations,
\begin{align}
{^t}N &= N, \label{Nsym} \\
B_{ab}{^t}N &= NB_{ab}, \label{NBab}
\end{align}
have to be satisfied.\footnote{See also Ref.~\cite{Kikuchi:2022psj}.}
The $N$ matrix which satisfies Eqs.~(\ref{Nsym}) and (\ref{NBab}) for $\forall a, b$ is just $N=n I$.
However, it is too restrictive for $N$ matrix, and then we consider the case with more relaxed $N$ matrix while the restricted modular symmetry than $\Gamma_{g}$.
In particular, to study non-trivial modular symmetry, we consider $N$ matrices which are consistent with $S$ and certain $T$ transformations generated by combinations of some of $T_{ab}$.
Thus, we consider the case that Eqs.~(\ref{Omega}), (\ref{Fflat}), and (\ref{Nsym}),~i.e.,
\begin{align}
&\Omega_{ij} = \Omega_{ji}, \label{Omegasymcomp} \\
&N_{ij} = N_{ji}, \label{Nsymcomp} \\
&\sum_{k} N_{ik}\Omega_{kj} = \sum_{k} \Omega_{ik}N_{kj}, \label{Fflatcomp} \\
\Leftrightarrow \ &\sum_{k \neq i,j} (N_{ik}\Omega_{jk} - N_{jk}\Omega_{ik}) + [ (N_{ii} - N_{jj}) \Omega_{ij} - N_{ij} (\Omega_{ii} - \Omega_{jj} ) ] = 0, \label{Fflatcompcond}
\end{align}
are satisfied.
Hereafter, we denote $\Delta N_{ij} \equiv N_{ii}-N_{jj}$ and $\Delta \Omega_{ij} \equiv \Omega_{ii}-\Omega_{jj}$.
Obviously, we find  $\Delta N_{ji}=-\Delta N_{ij}$ and $\Delta \Omega_{ji}=-\Delta \Omega_{ij}$.
Eq.~(\ref{Fflatcompcond}) is concretely written as
\begin{align}
\Delta N_{12} \Omega_{12} - N_{12} \Delta \Omega_{12} = 0, \label{Fflatcompcondg2}
\end{align}
for $g=2$ and
\begin{align}
&\left\{
\begin{array}{l}
(N_{31}\Omega_{23} - N_{23}\Omega_{31}) + (\Delta N_{12}\Omega_{12} - N_{12}\Delta \Omega_{12}) = 0 \\
(N_{12}\Omega_{31} - N_{31}\Omega_{12}) + (\Delta N_{23}\Omega_{23} - N_{23}\Delta \Omega_{23}) = 0 \\
(N_{12}\Omega_{23} - N_{23}\Omega_{12}) + (\Delta N_{31}\Omega_{31} - N_{31}\Delta \Omega_{31}) = 0
\end{array}
\right., \\
\Leftrightarrow&
\begin{pmatrix}
\Delta N_{12} & N_{31} & -N_{23} \\
-N_{31} & \Delta  N_{23} & N_{12} \\
N_{23} & -N_{12} & \Delta N_{31}
\end{pmatrix}
\begin{pmatrix}
\Omega_{12} \\ \Omega_{23} \\ \Omega_{31}
\end{pmatrix}
-
\begin{pmatrix}
N_{12} & & \\
 & N_{23} & \\
 & & N_{31}
\end{pmatrix}
\begin{pmatrix}
\Delta \Omega_{12} \\ \Delta \Omega_{23} \\ \Delta \Omega_{31}
\end{pmatrix}
=\begin{pmatrix}
0 \\ 0 \\ 0
\end{pmatrix},
\label{Fflatcompcondg3}
\end{align}
for $g=3$.
In the case of $g=1$, obviously, any moduli $\Omega$ are consistent with an arbitrary magnetic flux (matrix) $N$.
In the following, we classify the modular symmetry, which is consistent with a magnetic flux matrix $N$, for $g=1,2,3$.

First, let us see the case of $g=1$.
\begin{itemize}
\item[\bf{Class (1-1-a)}] There is only one case for $g=1$. Since any moduli $\Omega=\tau$ are consistent with a magnetic flux (matrix) $N=n$, we can consider $S$ and $T_{(11)}$ transformations, which generate $\Gamma_{1}=Sp(2,\mathbb{Z}) = SL(2,\mathbb{Z})$ transformation.
In particular, between $S$ and $T_{(11)}$, the following relation,
\begin{align}
(ST_{(11)})^3=I_2,
\end{align}
is satisfied.
The stabilizer is $H=\{\pm I_{2}\}=\mathbb{Z}_{2}^{t}$, which act on $z=z^1$ as
\begin{align}
\pm I_{2} : z^1 \rightarrow \pm z^1. \label{pmIz1}
\end{align}
Hence, $T^{2}/\mathbb{Z}_{2}^{t}$ twisted orbifold also has the same modular symmetry.
\end{itemize}

Next, let us see the case of $g=2$ in the following classes.
\begin{itemize}
\item[\bf{Class (2-1)}] In this class, we consider the case with $N_{12}=0$. Eq.~(\ref{Fflatcompcondg2}) is written by
\begin{align}
\Delta N_{12} \Omega_{12} = 0, \label{eq21}
\end{align}
and it is satisfied for $\forall \Delta \Omega_{12}$.
This class is further classified as follows.
\begin{itemize}
\item[\bf{Class (2-1-a)}] In this class, we also consider the case with $\Delta N_{12}=0$,~i.e.,
\begin{align}
N=
\begin{pmatrix}
n & 0 \\
0 & n
\end{pmatrix}
=nI_{2}.
\end{align}
Eq.~(\ref{eq21}) is also satisfied for $\forall \Omega_{12}$,~i.e.
\begin{align}
\Omega&=
\begin{pmatrix}
\tau_{11} & \tau_{12} \\
\tau_{12} & \tau_{22}
\end{pmatrix} \\
&= \sum_{i,j=1,2} \tau_{ij} B_{ij}. \notag
\end{align}
Then, we can consider $S$ and $T_{ab} \ (\forall a,b)$ transformations, which generate $\Gamma_{2}=Sp(4,\mathbb{Z})$ transformation.
In particular, among $S$ and $T_{ab}$, the following relations,
\begin{align}
\begin{array}{ll}
(ST_{11})^{6}=A_{B_{11}-B_{22}}, & (ST_{11})^{12}=I_{4}, \\
(ST_{22})^{6}=A_{B_{22}-B_{11}}, & (ST_{22})^{12}=I_{4}, \\
(ST_{12})^{3}=A_{B_{12}}, & (ST_{12})^{6}=I_{4},
\end{array}
\end{align}
are satisfied.
The stabilizer is $H=\{\pm I_{4}\}=\mathbb{Z}_2^{t}$, which act on $z={^t}(z^1, z^2)$ as
\begin{align}
\pm I_{4} :
\begin{pmatrix}
z^1 \\ z^2
\end{pmatrix}
\rightarrow
\begin{pmatrix}
\pm z^1 \\ \pm z^2
\end{pmatrix}.
\label{pmIz2}
\end{align}
Hence, $T^{4}/\mathbb{Z}_{2}^{t}$ twisted orbifold also has the same modular symmetry.
%
\item[\bf{Class (2-1-b)}] In this class, we consider the case with $\Delta N_{12} \neq 0$,~i.e.,
\begin{align}
N=
\begin{pmatrix}
n_{11} & 0 \\
0 & n_{22}
\end{pmatrix}.
\end{align}
Eq.~(\ref{eq21}) is satisfied for $\Omega_{12}=0$,~i.e.,
\begin{align}
\Omega&=
\begin{pmatrix}
\tau_{11} & 0 \\
0 & \tau_{22}
\end{pmatrix} \\
&= \sum_{i=1,2} \tau_{ii} B_{ii}. \notag
\end{align}
This is nothing but direct products of magnetized $T^2$ compactification.
Then, we can consider $S_{kk}$ and $T_{kk}$ $(k=1,2)$ transformations, which generate $N_{2}^{(2-1-b)}(H) =\otimes_{k=1}^{2} \Gamma_{1_k} = \otimes_{k=1}^{2} SL(2,\mathbb{Z})_k$ transformation.
Between $S_{kk}$ and $T_{kk}$, the following relation,
\begin{align}
(S_{kk}T_{kk})^3=I_{2},
\end{align}
is satisfied.
In particular, the stabilizer is $H=\{\pm I_{4}, \pm A_{B_{11}-B_{22}}\}=\otimes_{k=1}^{2} \mathbb{Z}_{2}^{t_k}$, which act on $z={^t}(z^1, z^2)$ as
\begin{align}
\pm A_{B_{11}-B_{22}}:
\begin{pmatrix}
z^1 \\ z^2
\end{pmatrix}
\rightarrow
\begin{pmatrix}
\pm z^1 \\ \mp z^2
\end{pmatrix},
\label{pmA1122z2}
\end{align}
in addition to Eq.~(\ref{pmIz2}).
Hence, $\otimes_{k=1}^{2} T^2_k/\mathbb{Z}_{2}^{t_k}$ twisted orbifold also has the same modular symmetry.
%
\end{itemize}
\item[\bf{Class (2-2)}] In this class, we consider the case with $N_{12} \neq 0$. Eq.~(\ref{Fflatcompcondg2}) is written by
\begin{align}
\Delta N_{12} \Omega_{12} = N_{12} \Delta \Omega_{12}. \label{eq22}
\end{align}
This class is further classified as follows.
\begin{itemize}
\item[\bf{Class (2-2-a)}] In this class, we also consider the case with $\Delta N_{12}=0$,~i.e.,
\begin{align}
N=
\begin{pmatrix}
n & n_{12} \\
n_{12} & n
\end{pmatrix}.
\end{align}
Eq.~(\ref{eq22}) is satisfied for $\Delta \Omega_{12}=0$ and $\forall \Omega_{12}$,~i.e.,
\begin{align}
\Omega&=
\begin{pmatrix}
\tau & \tau_{12} \\
\tau_{12} & \tau
\end{pmatrix} \\
&= \tau \sum_{i=1,2} B_{ii} + \tau_{12}B_{12} \notag \\
&\equiv \tau B_{I_2}+\tau_{12}B_{12}. \notag
\end{align}
Then, we can consider $S$, $T_{12}$, and $T_{I_2}$ with $B=B_{I_2}$ transformations, which generate $N_{2}^{(2-2-a)}(H)$ transformation.
In particular, between $S$ and $T$ transformation, the following relation,
\begin{align}
(ST_{I_2})^3=I_{4},
\end{align}
is satisfied.
In particular, the stabilizer is $H=\{\pm I_{4}, \pm A_{B_{12}} \}=\mathbb{Z}_{2}^{t} \times \mathbb{Z}_{2}^{p}$, which act on $z={^t}(z^1, z^2)$ as
\begin{align}
\pm A_{B_{12}} :
\begin{pmatrix}
z^1 \\ z^2
\end{pmatrix}
\rightarrow
\begin{pmatrix}
\pm z^2 \\ \pm z^1
\end{pmatrix},
\label{pmA12z2}
\end{align}
in addition to Eq.~(\ref{pmIz2}).
Hence, $T^{4}/(\mathbb{Z}_{2}^{t} \times \mathbb{Z}_{2}^{p})$ twisted and permutation orbifold also has the same modular symmetry.
%
\item[\bf{Class (2-2-b)}] In this class, we consider the case with $\Delta N_{12} \neq 0$,~i.e.,
\begin{align}
N=
\begin{pmatrix}
n_{11} & n_{12} \\
n_{12} & n_{22}
\end{pmatrix}.
\end{align}
Eq.~(\ref{eq22}) is satisfied if the following condition,
\begin{align}
\Omega_{12} : N_{12} = \Delta \Omega_{12} : \Delta N_{12}, \label{Omega12ratio}
\end{align}
is satisfied,~i.e.,
\begin{align}
\Omega&=
\begin{pmatrix}
\tau+\tau_{N}\Delta N_{12}/p & \tau_{N}n_{12}/p \\
\tau_{N}n_{12}/p & \tau
\end{pmatrix} \\
&= \tau \sum_{i=1,2} B_{ii} + \tau_{N}(\Delta N_{12}/p B_{11} + N_{12}/p B_{12}) \notag \\
&\equiv \tau B_{I_2}+\tau_{N} B_{N_2}, \notag
\end{align}
where $p={\rm gcd}(\Delta N_{12}, N_{12})$.
Note that the classes (2-1-b) and (2-2-a) are specific cases of the class (2-2-b).
Then, we can consider $S$, $T$, and 
$T_{N_2}$ with $B=B_{N_2} $transformations, which generate $N_{2}^{(2-2-b)}(H)$ transformation.
In particular, when $\Delta N_{12}=N_{12}$, between $S$ and $T_{N_2}$, the following relation,
\begin{align}
(ST_{N_2})^5=I_{4},
\end{align}
is satisfied, as shown in Appendix~\ref{ap:STn}.
The stabilizer is $H=\{\pm I_{4}\}=\mathbb{Z}_{2}^{t}$, which act on $z={^t}(z^1, z^2)$ as Eq.~(\ref{pmIz2}).
Hence, $T^{4}/\mathbb{Z}_{2}^{t}$ twisted orbifold also has the same modular symmetry.
\end{itemize}
\end{itemize}

Finally, let us see the case of $g=3$ in the following classes.
\begin{itemize}
\item[\bf{Class (3-1)}] In this class, we consider the case with $N_{12}=N_{23}=N_{31}=0$. Eq.~(\ref{Fflatcompcondg3}) is written by
\begin{align}
\begin{pmatrix}
\Delta N_{12} & 0 & 0 \\
0 & \Delta N_{23} & 0 \\
0 & 0 & \Delta N_{31}
\end{pmatrix}
\begin{pmatrix}
\Omega_{12} \\ \Omega_{23} \\ \Omega_{31}
\end{pmatrix}
=
\begin{pmatrix}
0 \\ 0 \\ 0
\end{pmatrix},
\label{eq31}
\end{align}
and it is satisfied for $\forall \Delta \Omega_{12}$, $\forall \Omega_{23}$, and $\forall \Omega_{31}$.
This class is further classified as follows.
\begin{itemize}
\item[\bf{Class (3-1-a)}] In this class, we also consider the case with $\Delta N_{12} = \Delta N_{23} = \Delta N_{31} =0$,~i.e.,
\begin{align}
N=
\begin{pmatrix}
n & 0 & 0 \\
0 & n & 0 \\
0 & 0 & n
\end{pmatrix}
=nI_{3}.
\end{align}
Eq.~(\ref{eq31}) is satisfied for $\forall \Omega_{12}$, $\forall \Omega_{23}$, and $\forall \Omega_{31}$,~i.e.,
\begin{align}
\Omega&=
\begin{pmatrix}
\tau_{11} & \tau_{12} & \tau_{13} \\
\tau_{12} & \tau_{22} & \tau_{23} \\
\tau_{13} & \tau_{23} & \tau_{33}
\end{pmatrix} \\
&= \sum_{i,j=1,2,3} \tau_{ij} B_{ij}. \notag
\end{align}
Then, we can consider $S$ and $T_{ab} \ (\forall a,b)$ transformations, which generate $\Gamma_{3}=Sp(6,\mathbb{Z})$ transformation.
In particular, among $S$ and $T_{ab}$, the following relations,
\begin{align}
\begin{array}{ll}
(ST_{11})^{6}=A_{B_{11}-B_{22}-B_{33}}, & (ST_{11})^{12}=I_{6}, \\
(ST_{22})^{6}=A_{B_{22}-B_{33}-B_{11}}, & (ST_{22})^{12}=I_{6}, \\
(ST_{33})^{6}=A_{B_{33}-B_{11}-B_{22}}, & (ST_{33})^{12}=I_{6}, \\
(ST_{12})^{6}=A_{B_{11}+B_{22}-B_{33}}, & (ST_{12})^{12}=I_{6}, \\
(ST_{23})^{6}=A_{B_{22}+B_{33}-B_{11}}, & (ST_{23})^{12}=I_{6}, \\
(ST_{31})^{6}=A_{B_{33}+B_{11}-B_{22}}, & (ST_{31})^{12}=I_{6},
\end{array}
\end{align}
are satisfied.
The stabilizer is $H=\{\pm I_{6}\}=\mathbb{Z}_2^{t}$, which act on $z={^t}(z^1, z^2, z^3)$ as
\begin{align}
\pm I_{6} :
\begin{pmatrix}
z^1 \\ z^2 \\ z^3
\end{pmatrix}
\rightarrow
\begin{pmatrix}
\pm z^1 \\ \pm z^2 \\ \pm z^3
\end{pmatrix}.
\label{pmIz3}
\end{align}
Hence, $T^{6}/\mathbb{Z}_{2}^{t}$ twisted orbifold also has the same modular symmetry.
%
\item[\bf{Class (3-1-b)}] In this class, we consider the case with $\Delta N_{23} \neq 0$, $\Delta N_{31} \neq 0$, and also $\Delta N_{12}=0$,~i.e.,
\begin{align}
N=
\begin{pmatrix}
n & 0 & 0 \\
0 & n & 0 \\
0 & 0 & n_{33}
\end{pmatrix}.
\end{align}
Eq.~(\ref{eq31}) is satisfied for $\Omega_{23}=\Omega_{31}=0$ and $\forall \Omega_{12}$,~i.e.,
\begin{align}
\Omega&=
\begin{pmatrix}
\tau_{11} & \tau_{12} & 0 \\
\tau_{12} & \tau_{22} & 0 \\
0 & 0 & \tau_{33}
\end{pmatrix} \\
&=\sum_{i,j=1,2} \tau_{ij}B_{ij} + \tau_{33}B_{33}. \notag
\end{align}
This is nothing but direct products of magnetized $T^4$ with the class (2-1-a) and $T^2$ compactifications.
Thus, we can consider $N_{3}^{(3-1-b)}(H)=\Gamma_{2} \times \Gamma_{1}=Sp(4,\mathbb{Z}) \times SL(2,\mathbb{Z})$ transformation.
$T^4/\mathbb{Z}_{2}^{t} \times T^2/\mathbb{Z}_{2}^{t}$ twisted orbifold also has the same modular symmetry.
On the other hand, when we have $\Delta N_{12} \neq 0$, $\Delta N_{23} \neq 0$, and also $\Delta N_{31} \neq 0$,~i.e.,
\begin{align}
N=
\begin{pmatrix}
n_{11} & 0 & 0 \\
0 & n_{22} & 0 \\
0 & 0 & n_{33}
\end{pmatrix},
\end{align}
Eq.~(\ref{eq31}) is satisfied for $\Omega_{12}=\Omega_{23}=\Omega_{31}=0$,~i.e.,
\begin{align}
\Omega&=
\begin{pmatrix}
\tau_{11} & 0 & 0 \\
0 & \tau_{22} & 0 \\
0 & 0 & \tau_{33}
\end{pmatrix} \\
&=\sum_{i=1,2,3} \tau_{ii}B_{ii} + \tau_{33}B_{33}. \notag
\end{align}
This is nothing but direct products of magnetized $T^4=T^2 \times T^2$ with the class (2-1-b) and $T^2$ compactifications.
Thus, we can consider $N_{3}^{(3-1-c)}(H)=\otimes_{k=1}^{3} \Gamma_{1_k} = \otimes_{k=1}^{3} SL(2,\mathbb{Z})_k$ transformation.
$\otimes_{k=1}^{3} T^2_k/\mathbb{Z}_{2}^{t_k}$ twisted orbifold also has the same modular symmetry.
%
\end{itemize}
\item[\bf{Class (3-2)}] In this class, we consider the case with $N_{23}=N_{31}=0$ and $N_{12} \neq 0$. Eq.~(\ref{Fflatcompcondg3}) is written by
\begin{align}
\begin{pmatrix}
\Delta N_{12} & 0 & 0 \\
0 & \Delta N_{23} & N_{12} \\
0 & -N_{12} & \Delta N_{31}
\end{pmatrix}
\begin{pmatrix}
\Omega_{12} \\ \Omega_{23} \\ \Omega_{31}
\end{pmatrix}
=
\begin{pmatrix}
N_{12} \Delta \Omega_{12} \\ 0 \\ 0
\end{pmatrix},
\label{eq32}
\end{align}
and it is satisfied for $\forall \Delta \Omega_{23}$ and $\forall \Delta \Omega_{31}$.
This class is further classified as follows.
\begin{itemize}
\item[\bf{Class (3-2-a)}] In this class, we also consider the case satisfying
\begin{align}
&\Delta N_{23} \Delta N_{31}+N_{12}^2 \neq 0, \notag \\
\Leftrightarrow \ &
N_{33} \neq \frac{(N_{11}+N_{22}) \pm \sqrt{\Delta N_{12}^2+(2N_{12})^2}}{2}, \label{N33neqEV}
\end{align}
that is,
\begin{align}
N=
\begin{pmatrix}
n_{11} & n_{12} & 0 \\
n_{12} & n_{22} & 0 \\
0 & 0 & n_{33}
\end{pmatrix}, \quad
n_{33} \neq \frac{(n_{11}+n_{22}) \pm \sqrt{(n_{11}-n_{22})^2+(2n_{12})^2}}{2}.
\end{align}
In particular, when $\Delta N_{12}=0 \Leftrightarrow n_{11}=n_{22}=n$ and $n_{33} \neq n \pm n_{12}$, 
Eq.~(\ref{eq32}) is satisfied for $\Delta \Omega_{12}=\Omega_{23}=\Omega_{31}=0$ and $\forall \Omega_{12}$,~i.e.,
\begin{align}
\Omega&=
\begin{pmatrix}
\tau & \tau_{12} & 0 \\
\tau_{12} & \tau & 0 \\
0 & 0 & \tau_{33}
\end{pmatrix} \\
&=\tau\sum_{i=1,2}B_{ii}+\tau_{12}B_{12}+\tau_{33}B_{33} \notag \\
&=\tau B_{I_2}+\tau_{12}B_{12}+\tau_{33}B_{33}. \notag
\end{align}
This is nothing but products of magnetized $T^4$ with the class (2-2-a) and $T^2$ compactifications.
Thus, we can consider $N_{3}^{(3-2-a)}(H)=N_{2}^{(2-2-a)}(H) \times \Gamma_{1}$ transformation.
$T^{4}/(\mathbb{Z}_{2}^{t} \times \mathbb{Z}_{2}^{p}) \times T^2/\mathbb{Z}_{2}^{t}$ orbifold also has the same modular symmetry.
On the other hand, when $\Delta N_{12} \neq 0$,
Eq.~(\ref{eq32}) is satisfied if $\Omega_{23}=\Omega_{31}=0$ and Eq.~(\ref{Omega12ratio}) are satisfied,~i.e.,
\begin{align}
\Omega&=
\begin{pmatrix}
\tau+\tau_{N}\Delta N_{12}/p & \tau_{N}N_{12}/p & 0 \\
\tau_{N}N_{12}/p & \tau & 0 \\
0 & 0 & \tau_{33}
\end{pmatrix} \\
&=\tau B_{I_2}+\tau_{N}B_{N_2}+\tau_{33}B_{33}. \notag
\end{align}
This is nothing but products of magnetized $T^4$ with the class (2-2-b) and $T^2$ compactifications.
Thus, we can consider $N_{3}^{(3-2-b)}(H)=N_{2}^{(2-2-b)}(H) \times \Gamma_{1}$ transformation.
$T^{4}/\mathbb{Z}_{2}^{t} \times T^2/\mathbb{Z}_{2}^{t}$ twisted orbifold also has the same modular symmetry.
%
\item[\bf{Class (3-2-b)}] On the other hand, in this class, we consider the case satisfying
\begin{align}
&\Delta N_{23} \Delta N_{31}+N_{12}^2 = 0, \notag \\
\Leftrightarrow \ &
N_{33} = \frac{(N_{11}+N_{22}) \pm \sqrt{\Delta N_{12}^2+(2N_{12})^2}}{2}, \label{N33neqEV}
\end{align}
that is,
\begin{align}
N=
\begin{pmatrix}
n_{11} & n_{12} & 0 \\
n_{12} & n_{22} & 0 \\
0 & 0 & n_{33}
\end{pmatrix}, \quad
n_{33} = \frac{(n_{11}+n_{22}) \pm \sqrt{(n_{11}-n_{22})^2+(2n_{12})^2}}{2} \in \mathbb{Z}.
\end{align}
In particular, when $\Delta N_{12}=0 \Leftrightarrow n_{11}=n_{22}=n$ and $n_{33} = n \pm n_{12}$, 
Eq.~(\ref{eq32}) is satisfied for $\Delta \Omega_{12}=0$, $\forall \Omega_{12}$, and $\Omega_{31} = \pm \Omega_{23}$,~i.e.,
\begin{align}
\Omega&=
\begin{pmatrix}
\tau & \tau_{12} & \tau' \\
\tau_{12} & \tau & \pm \tau' \\
\tau' & \pm \tau' & \tau_{33}
\end{pmatrix} \\
&=\tau B_{I_2}+\tau_{12}B_{12}+\tau_{33}B_{33}+\tau'(B_{13} \pm B_{23}), \notag \\
&\equiv \tau B_{I_2} + \tau_{12} B_{12} + \tau_{33} B_{33} + \tau' B'_{\pm} . \notag
\end{align}
On the other hand, when $\Delta N_{12} \neq 0$,
Eq.~(\ref{eq32}) is satisfied if Eq.~(\ref{Omega12ratio}) and
\begin{align}
\Omega_{23} : \Omega_{31} = N_{12} : \Delta N_{32} = \Delta N_{31} : N_{12} = r_{23}^{\pm} : r_{13}^{\pm} \quad ({\rm gcd}(r_{23}:r_{13})=1),
\end{align}
are satisfied,~i.e.,
\begin{align}
\Omega&=
\begin{pmatrix}
\tau+\tau_{N}\Delta N_{12}/p & \tau_{N}N_{12}/p & \tau' r_{13}^{\pm} \\
\tau_{N}N_{12}/p & \tau & \pm \tau' r_{23}^{\pm} \\
\tau' r_{13}^{\pm} & \tau' r_{23}^{\pm} & \tau_{33}
\end{pmatrix} \\
&=\tau B_{I_2}+\tau_{N}B_{N_2}+\tau_{33}B_{33}+\tau' \sum_{i=1,2} r_{i3}^{\pm} B_{i3}. \notag \\
&\equiv \tau B_{I_2}+\tau_{N}B_{N_2}+\tau_{33}B_{33}+\tau' B'_{\pm}. \notag
\end{align}
Note that the former case is the specific case of the latter case.
Then, we can consider $T'_{\pm}$ with $B'_{\pm}$ transformation in addition to transformations by $N_{3}^{(3-2-a)}$, which generate $N_{3}^{(3-2-b)}(H)$ transformation.
In particular, when $n_{11}=n_{22}=n$ and $n_{33}=n \pm n_{12}$, between $S$ and $T'_{\pm}$, the following relations,
\begin{align}
(ST'_{\pm})^4=A_{\mp B_{12}-B_{33}}, \quad (ST'_{\pm})^8=I_{6},
\end{align}
are satisfied, as shown in Appendix~\ref{ap:STn}.
Except for the case with $n_{11}=n_{22}=n$ and $n_{33}=n+n_{12}$,
the stabilizer is $H=\{\pm I_{6}\}=\mathbb{Z}_{2}^{t}$, which act on $z={^t}(z^1, z^2, z^3)$ as
\begin{align}
\pm I_{6} :
\begin{pmatrix}
z^1 \\ z^2 \\ z^3
\end{pmatrix}
\rightarrow
\begin{pmatrix}
\pm z^1 \\ \pm z^2 \\ \pm z^3
\end{pmatrix}.
\label{pmIz3}
\end{align}
Hence, $T^{6}/\mathbb{Z}_{2}^{t}$ twisted orbifold also has the same modular symmetry.
When $n_{11}=n_{22}=n$ and $n_{33}=n+n_{12}$, on the other hand, the stabilizer is $H=\{\pm I_{6}, \pm A_{B_{12}+B_{33}} \}=\mathbb{Z}_{2}^{t} \times \mathbb{Z}_{2}^{p}$, which act on $z={^t}(z^1, z^2, z^3)$ as
\begin{align}
\pm A_{B_{12}+B_{33}} :
\begin{pmatrix}
z^1 \\ z^2 \\ z^3
\end{pmatrix}
\rightarrow
\begin{pmatrix}
\pm z^2 \\ \pm z^1 \\ \pm z^3
\end{pmatrix},
\label{pmA1233z3}
\end{align}
in addition to Eq.~(\ref{pmIz3}).
Hence, $T^{6}/(\mathbb{Z}_{2}^{t} \times \mathbb{Z}_{2}^{p})$ twisted and permutation orbifold also has the same modular symmetry.
%
\end{itemize}
\item[\bf{Class (3-3)}] In this class, we consider the case with $N_{12} \neq 0$, $N_{13} \neq 0$, and $N_{23}=0$ case,~i.e.,
\begin{align}
N=
\begin{pmatrix}
n_{11} & n_{12} & n_{13} \\
n_{12} & n_{22} & 0 \\
n_{13} & 0 & n_{33}
\end{pmatrix}.
\end{align}
Eq.~(\ref{Fflatcompcondg3}) is written by
\begin{align}
\begin{pmatrix}
\Delta N_{12} & N_{31} & 0 \\
-N_{31} & \Delta N_{23} & N_{12} \\
0 & -N_{12} & \Delta N_{31}
\end{pmatrix}
\begin{pmatrix}
\Omega_{12} \\ \Omega_{23} \\ \Omega_{31}
\end{pmatrix}
=
\begin{pmatrix}
N_{12} \Delta \Omega_{12} \\ 0 \\ N_{31} \Delta N_{31}
\end{pmatrix}.
\label{eq33}
\end{align}
Then, it is satisfied
when the following relations,
\begin{align}
\begin{array}{l}
\Omega_{12} = \tau_{N} N_{12}/p + \frac{\Delta N_{23}}{N_{31}} \Omega_{23} \\
\Omega_{31} = \tau_{N} N_{31}/p \\
\Delta \Omega_{12} = \frac{\Delta N_{12}}{N_{12}} \Omega_{12} + \frac{N_{31}}{N_{12}} \Omega_{23}, \\
\Delta \Omega_{31} = \frac{\Delta N_{31}}{N_{31}} \Omega_{31} - \frac{N_{12}}{N_{31}} \Omega_{23}, \\
\Delta \Omega_{23} = -\frac{\Delta N_{12}}{N_{12}} \Omega_{12} -\frac{\Delta N_{31}}{N_{31}} \Omega_{31} + \frac{N_{12}^2-N_{13}^2}{N_{12}N_{13}} \Omega_{23},
\end{array}
\end{align}
are satisfied,~i.e.,
\begin{align}
\Omega=&
\tau
\begin{pmatrix}
1 & 0 & 0 \\
0 & 1 & 0 \\
0 & 0 & 1
\end{pmatrix}
+\tau_{N}
\begin{pmatrix}
\Delta N_{12}/p & N_{12}/p & N_{13}/p \\
N_{12}/p & 0 & 0 \\
N_{13}/p & 0 & -\Delta N_{23}/p
\end{pmatrix} \notag \\
&+\tau_{N^{-1}}
\begin{pmatrix}
(\Delta N_{12} \Delta N_{23} + N_{13}^2)/p^2 & N_{12}\Delta N_{23}/p^2 & 0 \\
N_{12}\Delta N_{23}/p^2 & 0 & N_{12}N_{13}/p^2 \\
0 & N_{12}N_{13}/p^2 &  (\Delta N_{12} \Delta N_{23} + N_{13}^2 - N_{12}^2)/p^2
\end{pmatrix} \\
=&\tau\sum_{i=1,2,3}B_{ii}+\tau_{N}(\Delta N_{12}/p B_{11}-\Delta N_{31}/p B_{33}+\sum_{i=2,3}N_{1i}/pB_{1i}) \notag \\
&+\tau_{N^{-1}}[(\Delta N_{12} \Delta N_{23}+N_{13}^2)/p^2B_{11}-(\Delta N_{12} \Delta N_{23}+N_{13}^2-N_{12}^2)/p^2 B_{33} \notag \\
&\quad +\Delta N_{12} \Delta N_{23}/p^2+N_{12}N_{13}/p^2B_{23}], \notag \\
\equiv& \tau B_{I_3} + \tau_{N} B_{N_3} +\tau_{N^{-1}} B_{N_3^{-1}}, \notag
\end{align}
where $p={\rm gcd}(\Delta N_{12}, \Delta N_{23}, N_{12},N_{13})$.
Then, we can consider $S$, $T$, $T_{N_3}$ with $B=B_{N_3}$, and $T_{N_3^{-1}}$ with $B=B_{N_3^{-1}}$ transformations, which generate $N_{3}^{(3-3)}(H)$ transformation.
In particular, when $\Delta N_{12}=\Delta N_{23}=\Delta N_{31}=0$ and $N_{12}=N_{13}$, among $T_{I_3}$, $T_{N_3}$, $T_{N_3^{-1}}$, and $S$, the following relations,
\begin{align}
\begin{array}{ll}
(ST_{I_3})^3=I_{6}, \\
(ST_{N_3})^4=-A_{B_{11}+B_{23}}, & (ST_{N_3})^8=I_{6}, \\
(ST_{N_3^{-1}})^3=A_{B_{11}+B_{23}}, & (ST_{N_3^{-1}})^6=I_{6},
\end{array}
\end{align}
are satisfied, as shown in Appendix~\ref{ap:STn}.
Except for the case with $\Delta N_{23}=0$ and $N_{12}=N_{13}$,
the stabilizer is $H=\{ \pm I_{6} \}=\mathbb{Z}_{2}^{t}$, which act on $z={^t}(z^1, z^2, z^3)$ as Eq.~(\ref{pmIz3}).
Hence, $T^{6}/\mathbb{Z}_{2}^{t}$ twisted orbifold also has the same modular symmetry.
When $\Delta N_{23}=0$ and $N_{12}=N_{13}$, on the other hand, the stabilizer is $H=\{\pm I_{6}, \pm A_{B_{11}+B_{23}} \}=\mathbb{Z}_{2}^{t} \times \mathbb{Z}_{2}^{p}$, which act on $z={^t}(z^1, z^2, z^3)$ as
\begin{align}
\pm A_{B_{11}+B_{23}} :
\begin{pmatrix}
z^1 \\ z^2 \\ z^3
\end{pmatrix}
\rightarrow
\begin{pmatrix}
\pm z^1 \\ \pm z^3 \\ \pm z^2
\end{pmatrix},
\label{pmA1123z3}
\end{align}
in addition to Eq.~(\ref{pmIz3}).
Hence, $T^{6}/(\mathbb{Z}_{2}^{t} \times \mathbb{Z}_{2}^{p})$ twisted and permutation orbifold also has the same modular symmetry.
%
\item[\bf{Class (3-4)}] In this class, we consider the case with $N_{12} \neq  0$, $N_{23} \neq 0$, and $N_{31} \neq 0$. Eq.~(\ref{Fflatcompcondg3}) is written by
\begin{align}
\begin{pmatrix}
\Delta \Omega_{12} \\ \Delta \Omega_{23} \\ \Delta \Omega_{31}
\end{pmatrix}
&=
\begin{pmatrix}
N_{12}^{-1} & 0 & 0 \\
0 & N_{23}^{-1} & 0 \\
0 & 0 & N_{31}^{-1}
\end{pmatrix}
\begin{pmatrix}
\Delta N_{12} & N_{31} & -N_{23} \\
-N_{31} & \Delta N_{23} & N_{12} \\
N_{23} & -N_{12} & \Delta N_{31}
\end{pmatrix}
\begin{pmatrix}
\Omega_{12} \\ \Omega_{23} \\ \Omega_{31}
\end{pmatrix} \notag \\
&=
\begin{pmatrix}
\frac{\Delta N_{12}}{N_{12}} \Omega_{12} + \frac{N_{31}}{N_{12}} \Omega_{23} - \frac{N_{23}}{N_{12}} \Omega_{31} \\
-\frac{N_{31}}{N_{23}} \Omega_{12} + \frac{\Delta N_{23}}{N_{23}} \Omega_{23} + \frac{N_{12}}{N_{23}} \Omega_{31} \\
\frac{N_{23}}{N_{31}} \Omega_{12} - \frac{N_{12}}{N_{31}} \Omega_{23} + \frac{\Delta N_{31}}{N_{31}} \Omega_{31}
\end{pmatrix}.
\label{eq34}
\end{align}
Note that it is required that
\begin{align}
&\Delta \Omega_{12} + \Delta \Omega_{23} + \Delta \Omega_{31} = 0, \notag \\
\Leftrightarrow \ &
\begin{pmatrix}
\frac{\Delta N_{12}}{N_{12}} - \frac{N_{31}^2 - N_{23}^2}{N_{23}N_{31}} & \frac{\Delta N_{23}}{N_{23}} - \frac{N_{12}^2 - N_{31}^2}{N_{31}N_{12}} & \frac{\Delta N_{31}}{N_{31}} - \frac{N_{23}^2 - N_{12}^2}{N_{12}N_{23}}
\end{pmatrix}
\begin{pmatrix}
\Omega_{12} \\ \Omega_{23} \\ \Omega_{31}
\end{pmatrix}
=0. \label{orthogonalcond}
\end{align}
This class is further classified as follows.
\begin{itemize}
\item[\bf{Class (3-4-a)}] In this class, we consider the case that
\begin{align}
N=
\begin{pmatrix}
N_{11} & N_{12} & N_{31} \\
N_{12} & N_{22} & N_{23} \\
N_{31} & N_{23} & N_{33}
\end{pmatrix}, \quad
\left\{
\begin{array}{l}
N_{11} = n + \frac{N_{31}N_{12}}{N_{23}} \in \mathbb{Z}, \\
N_{22} = n + \frac{N_{12}N_{23}}{N_{31}} \in \mathbb{Z}, \\
N_{33} = n + \frac{N_{23}N_{31}}{N_{12}} \in \mathbb{Z},
\end{array}
\right..
\end{align}
Eq.~(\ref{orthogonalcond}) is satisfied for $\forall \Omega_{12}$, $\forall \Omega_{23}$, and $\forall \Omega_{31}$.
By combining Eq.~(\ref{eq34}), the moduli in this case are given by
\begin{align}
\Omega=&
\tau
\begin{pmatrix}
1 & 0 & 0 \\
0 & 1 & 0 \\
0 & 0 & 1
\end{pmatrix}
+\tau'_{12}
\begin{pmatrix}
(N_{31}^2-N_{23}^2)/p_{12}^2 & N_{23}N_{31}/p_{12}^2 & 0 \\
N_{23}N_{31}/p_{12}^2 & 0 & 0 \\
0 & 0 & N_{31}^2/p_{23}^2
\end{pmatrix} \notag \\
&+\tau'_{23}
\begin{pmatrix}
N_{12}^2/p_{23}^2 & 0 & 0 \\
0 & (N_{12}^2-N_{31}^2)/p_{23}^2 & N_{31}N_{12}/p_{23}^2 \\
0 & N_{31}N_{12}/p_{23}^2 & 0
\end{pmatrix} \notag \\
&+\tau'_{31}
\begin{pmatrix}
0 & 0 & N_{12}N_{23}/p_{31}^2 \\
0 & N_{23}^2/p_{31}^2 & 0 \\
N_{12}N_{23}/p_{31}^2 & 0 & (N_{23}^2-N_{12}^2)/p_{31}^2
\end{pmatrix} \\
=&\tau \sum_{i=1,2,3} B_{ii}+\tau'_{12}(N_{23}N_{31}/p_{12}^2 B_{12} + N_{31}^2/p_{12}^2 B_{33} + (N_{31}^2-N_{23}^2)/p_{12}^2 B_{11} ) \notag \\
&+\tau'_{23}(N_{31}N_{12}/p_{23}^2 B_{23} + N_{12}^2/p_{23}^2 B_{11} + (N_{12}^2-N_{31}^2)/p_{23}^2 B_{22} ) \notag \\
&+\tau'_{31}(N_{12}N_{23}/p_{31}^2 B_{13} + N_{23}^2/p_{31}^2 B_{22} + (N_{23}^2-N_{12}^2)/p_{31}^2 B_{33} ) \notag \\
\equiv& \tau B_{I_3} + \tau'_{12} B'_{12} + \tau'_{23} B_{23} + \tau'_{31} B'_{13},
\end{align}
where $p_{12}={\rm gcd}(N_{23},N_{31})$, $p_{23}={\rm gcd}(N_{31},N_{12})$, and $p_{31}={\rm gcd}(N_{12},N_{23})$.
Then, we can consider $S$, $T$, $T'_{12}$ with $B=B'_{12}$, $T'_{23}$ with $B=B'_{23}$, and $T'_{31}$ with $B=B'_{31}$, which generate $N_{3}^{(3-4-a)}(H)$ transformation.
In particular, when $N_{12}=N_{23}=N_{31}$, among $T'_{12}$, $T'_{23}$, $T'_{31}$, and $S$, the following relations,
\begin{align}
\begin{array}{ll}
(ST'_{12})^3=A_{B_{12}+B_{33}}, & (ST'_{12})^6=I_{6}, \\
(ST'_{23})^3=A_{B_{23}+B_{11}}, & (ST'_{23})^6=I_{6}, \\
(ST'_{31})^3=A_{B_{31}+B_{22}}, & (ST'_{31})^6=I_{6},
\end{array}
\end{align}
are satisfied.
The stabilizer is generally $H=\{ \pm I_{6} \}=\mathbb{Z}_{2}^{t}$, which act on $z={^t}(z^1, z^2, z^3)$ as Eq.~(\ref{pmIz3}).
Hence, $T^{6}/\mathbb{Z}_{2}^{t}$ twisted orbifold also has the same modular symmetry.
In particular, when $\tau'_{12}=\tau'_{23}=\tau'_{31}$ is also satisfied, the stabilizer becomes $H=\{ \pm I_{6}, \pm A_{B_{12}+B_{33}}, \pm A_{B_{23}+B_{11}}, \pm A_{B_{31}+B_{22}}, \pm A_{P_{231}}, \pm A_{P_{312}} \}=\mathbb{Z}_{2}^{t} \times S_3$, which act on $z={^t}(z^1, z^2, z^3)$ as
\begin{align}
\pm A_{P_{231}} :
\begin{pmatrix}
z^1 \\ z^2 \\ z^3
\end{pmatrix}
&\rightarrow
\begin{pmatrix}
\pm z^2 \\ \pm z^3 \\ \pm z^1
\end{pmatrix},
\label{pmA231z3} \\
\pm A_{P_{312}} :
\begin{pmatrix}
z^1 \\ z^2 \\ z^3
\end{pmatrix}
&\rightarrow
\begin{pmatrix}
\pm z^3 \\ \pm z^1 \\ \pm z^2
\end{pmatrix},
\label{pmA312z3} \\
\pm A_{B_{31}+B_{22}} :
\begin{pmatrix}
z^1 \\ z^2 \\ z^3
\end{pmatrix}
&\rightarrow
\begin{pmatrix}
\pm z^3 \\ \pm z^2 \\ \pm z^1
\end{pmatrix},
\label{pmA3122z3}
\end{align}
in addition to Eqs.~(\ref{pmA1233z3}), (\ref{pmA1123z3}), and (\ref{pmIz3}).
We note that
\begin{align}
\begin{array}{l}
A_{B_{12}+B_{33}} A_{B_{31}+B_{22}} = A_{B_{31}+B_{22}} A_{B_{23}+B_{11}} = A_{B_{23}+B_{11}} A_{B_{12}+B_{33}} = A_{P_{231}}, \\ 
A_{B_{31}+B_{22}} A_{B_{12}+B_{33}} = A_{B_{23}+B_{11}} A_{B_{31}+B_{22}} = A_{B_{12}+B_{33}} A_{B_{23}+B_{11}}= A_{P_{312}}.
\end{array}
\end{align}
Hence, $T^{6}/(\mathbb{Z}_{2}^{t} \times S_3)$ orbifold also has the same modular symmetry.
%
\item[\bf{Class (3-4-b)}] In this class, we consider the case with the vector,
\begin{align}
\begin{pmatrix}
\frac{\Delta N_{12}}{N_{12}} - \frac{N_{31}^2 - N_{23}^2}{N_{23}N_{31}} \\
\frac{\Delta N_{23}}{N_{23}} - \frac{N_{12}^2 - N_{31}^2}{N_{31}N_{12}} \\
\frac{\Delta N_{31}}{N_{31}} - \frac{N_{23}^2 - N_{12}^2}{N_{12}N_{23}}
\end{pmatrix}
\neq
\begin{pmatrix}
0 \\ 0 \\ 0
\end{pmatrix}.
\end{align}
In this case,
there are two degrees of freedom of the vector, ${^t}(\Omega_{12}, \Omega_{23}, \Omega_{31})$, which satisfying Eq.~(\ref{orthogonalcond}).
Indeed, that vector can be written as
\begin{align}
\begin{pmatrix}
\Omega_{12} \\ \Omega_{23} \\ \Omega_{31}
\end{pmatrix}
=\tau_{N}
\begin{pmatrix}
N_{12} \\ N_{23} \\ N_{31}
\end{pmatrix}
+\tau_{N^{-1}}
\begin{pmatrix}
N_{23}N_{31}-N_{12}N_{33} \\ N_{31}N_{12}-N_{23}N_{11} \\ N_{12}N_{23}-N_{31}N_{22}
\end{pmatrix},
\end{align}
and then, from Eq.~(\ref{eq34}), we can obtain that
\begin{align}
\begin{pmatrix}
\Delta \Omega_{12} \\ \Delta \Omega_{23} \\ \Delta \Omega_{31}
\end{pmatrix}
=\tau_{N}
\begin{pmatrix}
\Delta N_{12} \\ \Delta N_{23} \\ \Delta N_{31}
\end{pmatrix}
+\tau_{N^{-1}}
\begin{pmatrix}
N_{31}^2-N_{12}^2-\Delta N_{12}N_{33} \\ N_{12}^2-N_{31}^2-\Delta N_{23}N_{11} \\ N_{23}^2-N_{12}^2-\Delta N_{31}N_{22}
\end{pmatrix}.
\end{align}
Thus, the moduli can be written by
\begin{align}
\Omega=&
\tau
\begin{pmatrix}
1 & 0 & 0 \\
0 & 1 & 0 \\
0 & 0 & 1
\end{pmatrix}
+\tau_{N}/p_{N}
\begin{pmatrix}
N_{11} & N_{12} & N_{31} \\
N_{12} & N_{22} & N_{23} \\
N_{31} & N_{23} & N_{33}
\end{pmatrix} \notag \\
&+\tau_{N^{-1}}/p_{N^{-1}}
\begin{pmatrix}
N_{22}N_{33}-N_{23}^2 & N_{23}N_{31}-N_{12}N_{33} & N_{12}N_{23}-N_{31}N_{22} \\
N_{23}N_{31}-N_{12}N_{33} & N_{33}N_{11}-N_{31}^2 & N_{31}N_{12}-N_{23}N_{11} \\
N_{12}N_{23}-N_{31}N_{22} & N_{31}N_{12}-N_{23}N_{11} & N_{11}N_{22}-N_{12}^2
\end{pmatrix} \\
=& \tau \sum B_{ii} + \tau_{N} \sum N_{ij}/p B_{ij} + \tau_{N^{-1}} \sum \tilde{N}_{ij}/q B_{ij} \notag \\
\equiv& \tau B_{I_3} + \tau_{N} B_{N_3} + \tau_{N^{-1}} B_{N_3^{-1}}, \notag
\end{align}
where $p_{N}$ and $p_{N^{-1}}$ denote the greatest common divisor of all elements of the $N$ matrix and the adjugate matrix $\tilde{N}$ (defined by $N^{-1}=({\rm det}N)^{-1} \tilde{N}$), respectively.
Then, we can consider $S$, $T$, $T_{N_3}$ with $B=B_{N_3}$, and $T_{N_3^{-1}}$ with $B=B_{N_3^{-1}}$, which generate $N_{3}^{(3-4-b)}(H)$ transformation.
The stabilizer is $H=\{ \pm I_{6} \}=\mathbb{Z}_{2}^{t}$, which act on $z={^t}(z^1, z^2, z^3)$ as Eq.~(\ref{pmIz3}).
Hence, $T^{6}/\mathbb{Z}_{2}^{t}$ twisted orbifold also has the same modular symmetry.
%
\end{itemize}
\end{itemize}

In the above cases, not only the magnetic flux $F$ but the gauge potential $A$ and the covariant derivative $D$ as well as the Dirac operator $i\slash{D}$ are invariant under the corresponding modular transformation.


\subsection{Modular symmetry of wave functions on magnetized $T^{2g}$ and orbifold models}
\label{subsec:MSWMT2gOM}

Now, let us see the modular symmetry of wave functions on magnetized $T^{2g}$ in Eq.~(\ref{wavT2g}).
First, in order for the boundary conditions of wave functions on magnetized $T^{2g}$ in Eqs.~(\ref{BCPsiz1}) and (\ref{BCPsiztau}) to be consistent with the modular transformation, particularly for $S$ and $T$ transformations, the following conditions,
\begin{align}
S\ {\rm transformation}\ &\Rightarrow \ \beta^S_k=\alpha^S_k=0, \ {\rm or} \ 1/2, \label{Swavcond} \\
T\ {\rm transformation}\ &\Rightarrow \ (NB)_{ii} + (2{^t}\alpha^S B)_{i} = \sum_{k} (N_{ik} + 2\alpha^S_k) (B)_{ki} \in 2\mathbb{Z}, \label{Twavcond}
\end{align}
are required,\footnote{When $B=B_{I_{g}}$, it is consistent with the analysis in Ref.~\cite{Kikuchi:2021ogn}.} in addition to Eqs.~(\ref{Omegasymcomp})-(\ref{Fflatcompcond}), where $B$ denotes a $g$ dimensional symmetric matrix generated by some combinations of $B_{ab}$.
In particular, in order to consider full $\Gamma_{g}=Sp(2g,\mathbb{Z})$ transformation for wave functions on magnetized $T^{2g}$, it is required that the $N$ matrix is in the class ($g$-1-a) with $n \in 2\mathbb{Z}$ and the SS phases are $\alpha^S_k=\beta^S_k=0$.
Hereafter, we consider vanishing SS phases, and then we omit the SS phase indices.
In this case, in order to satisfy Eq.~(\ref{Twavcond}), the $N$ matrix in each class must be $\forall N_{ij} \in 2\mathbb{Z}$.
In the following, we often denote the $N$ matrix as,
\begin{align}
N = s N',
\end{align}
where $s$ is the greatest common divisor of all components of the $N$ matrix, $\forall N_{ij}$.

When the above conditions including conditions in individual classes discussed in the previous subsection, are satisfied, wave functions on magnetized $T^{2g}$ in Eq.~(\ref{wavT2g}) transform under the modular transformation~\cite{Mumford:1984,Kikuchi:2022psj}, as
\begin{align}
&\widetilde{\gamma} : \psi^{J,N}_{T^{2g}}(z,\Omega) \rightarrow \psi^{J,N}_{T^{2g}}(\widetilde{\gamma}(z,\Omega)) = \widetilde{J}_{1/2}(\widetilde{\gamma},\Omega) \rho_{T^{2g}}(\widetilde{\gamma})_{JK} \psi^{K,N}_{T^{2g}}(z,\Omega), \label{modularwave} \\
&\widetilde{S} = [S,(-1)^g] : \quad \widetilde{J}_{1/2}(\widetilde{S},\Omega) = (-1)^g ({\rm det}(-\Omega))^{1/2}, \quad \rho_{T^{2g}}(\widetilde{S})_{JK} = \frac{(-e^{\pi i/4})^g}{\sqrt{{\rm det}N}} e^{2\pi i{^t}JN^{-1}K}, \label{Stilde} \\
&\widetilde{T} = [T,1] :  \quad \widetilde{J}_{1/2}(\widetilde{T},\Omega) = 1, \quad \rho_{T^{2g}}(\widetilde{T})_{JK} = e^{\pi i{^t}JBN^{-1}J} \delta_{J,K}. \label{Ttilde}
\end{align}
This means that the wave functions behave as the Siegel modular forms of weight $1/2$.
Then, as mentioned in section~\ref{sec:MT2g}, the corresponding 4D chiral fields, $\varphi^{J}(x)$, also transform as
\begin{align}
\widetilde{\gamma} : \varphi^{J}(x) \rightarrow \widetilde{J}_{-1/2}(\widetilde{\gamma},\Omega) \bar{\rho}_{T^{2g}}(\widetilde{\gamma})_{JK} \varphi^{K}(x).
\end{align}
This means that the 4D chiral fields transform non-trivially under $\bar{\rho}_{T^{2g}}(\widetilde{\gamma})$ modular flavor transformation with modular weight $-1/2$,~\footnote{That is, the sign of the modular weight of 4D fields is flipped from one in extra dimensions.} which can be also found from Eq.~(\ref{Normalization}).
In the following, we study under what modular flavor group the wave functions transform non-trivially.
Here, we note that the following relation~\cite{Mumford:1984,Kikuchi:2022psj} should, in particular, be satisfied,
\begin{align}
\begin{array}{l}
\psi^{J,N}_{T^{2g}}(\widetilde{A}_{X}^{\pm}(z,\Omega)) = \widetilde{J}_{1/2}(\widetilde{A}_{X}^{\pm}, \Omega) \rho_{T^{2g}}(\widetilde{A}_{X}^{\pm})_{JK} \psi^{K,N}_{T^{2g}}(z,\Omega) = \psi^{XJ,N}_{T^{2g}}(z,\Omega), \\
\widetilde{A}_{X}^{\pm} \equiv [A_{X},\pm 1] : \ \widetilde{J}_{1/2}(\widetilde{A}_{X}^{\pm},\Omega) = \pm ({\rm det}X)^{1/2}, \ \rho_{T^{2g}}(\widetilde{A}_{X}^{\pm})_{JK} = \pm ({\rm det}X)^{-1/2} \delta_{XJ,K}.
\end{array}
\label{Atildewave}
\end{align}
Indeed, as for $S^2=A_{-I_{g}}=-I_{2g}$ transformation, we can find that
\begin{align}
\begin{array}{l}
\rho_{T^{2g}}(\widetilde{R})_{JK} = \rho_{T^{2g}}(\widetilde{S})_{JK}^2 = e^{\pi ig/2} \delta_{-J,K}. \\
\rho_{T^{2g}}(\widetilde{R})_{JK}^2 = \rho_{T^{2g}}(\widetilde{S})_{JK}^4 = (-1)^g \delta_{J,K}, \\
\rho_{T^{2g}}(\widetilde{R})_{JK}^4 = \rho_{T^{2g}}(\widetilde{S})_{JK}^8 =\delta_{J,K}.
\end{array}
\label{Stildewaverel}
\end{align}
As for $(ST)^n=A_{\Omega_0^n}$ transformation, it depends on the detail structure of $B$ matrix in $T$ transformation, as in Appendix~\ref{ap:STn}.
For example, when $B^2=I_{g}$ is satisfied, we can find that
\begin{align}
\begin{array}{l}
\ [ \rho_{T^{2g}}(\widetilde{S}) \rho_{T^{2g}}(\widetilde{T}) ]^3_{JK} = e^{-\pi i n^{B}_{-}/2} \delta_{BJ,K}, \\
\ [ \rho_{T^{2g}}(\widetilde{S}) \rho_{T^{2g}}(\widetilde{T}) ]^6_{JK} = (-1)^{n^{B}_{-}} \delta_{J,K}, \\
\ [ \rho_{T^{2g}}(\widetilde{S}) \rho_{T^{2g}}(\widetilde{T}) ]^{12}_{JK} = \delta_{J,K},
\end{array}
\label{ST3tildewaverel}
\end{align}
where $n^{B}_{-}$ denotes the number of negative eigenvalues of $B$ matrix and we apply the $g$-dimensional Landsberg-Schaar relation,
\begin{align}
\frac{1}{\sqrt{{\rm det}N}} \sum_{K \in \Lambda_N} e^{\pi i {^t}K N^{-1}B K} = \frac{e^{\pi i(g-2n^{B}_{-})/4}}{\sqrt{|{\rm det}B}|} \sum_{K \in \Lambda_B} e^{-\pi i {^t}K NB^{-1} K}, \label{LSrel}
\end{align}
for the above calculations.
(See for the Landsberg-Schaar relation Appendix \ref{ap:Landsberg-Schaar}.)
By considering Eqs.~(\ref{Stildewaverel}) and (\ref{ST3tildewaverel}), we can also find that
\begin{align}
\begin{array}{l}
\ [ \rho_{T^{2g}}(\widetilde{S}) \rho_{T^{2g}}(\widetilde{T}) ]^6_{JK} = (-1)^{g'-1} e^{3\pi i(g-g')/2} \delta_{(-I_{g}+2B_{ab}^2)J,K}, \\
\ [ \rho_{T^{2g}}(\widetilde{S}) \rho_{T^{2g}}(\widetilde{T}) ]^{12}_{JK} = (-1)^{g-g'} \delta_{J,K}, \\
\ [ \rho_{T^{2g}}(\widetilde{S}) \rho_{T^{2g}}(\widetilde{T}) ]^{24}_{JK} = \delta_{J,K}.
\end{array}
\label{STab6tildewaverel}
\end{align}
Similarly, by utilizing the relation in Eq.~(\ref{LSrel}), we can calculate the other relations among $S$ and $T$,~e.g.,
\begin{align}
&\begin{array}{l}
\ [ \rho_{T^{4}}(\widetilde{S}) \rho_{T^{4}}(\widetilde{T}) ]_{JK}^5 = - \delta_{J,K}, \quad (T=T_{aa}T_{12}\ (a=1,2))\\
\ [ \rho_{T^{4}}(\widetilde{S}) \rho_{T^{4}}(\widetilde{T}) ]_{JK}^{10} = \delta_{J,K},
\end{array}
\label{STN5tildewaverel} \\
&\begin{array}{l}
\ [ \rho_{T^{4}}(\widetilde{S}) \rho_{T^{4}}(\widetilde{T}) ]_{JK}^4 = - \delta_{\Omega_0^4 J,K}, \quad (T=T_{ab}T_{bc}\ ((a,b,c)=(1,2,3),(2,3,1),(3,1,2))) \\
\ [ \rho_{T^{4}}(\widetilde{S}) \rho_{T^{4}}(\widetilde{T}) ]_{JK}^{8} = \delta_{J,K}.
\end{array}
\label{ST4tildewaverel} 
\end{align}
In addition, we obtain the following relations among $T$ transformations,
\begin{align}
\begin{array}{l}
\rho_{T^{2g}}(\widetilde{T}_{X})_{JK}^{h_{X}} = \delta_{J,K}, \ 
h_{X} = \left\{
\begin{array}{ll}
s{\rm det}N' & ((B_{X}\tilde{N}')_{ii} \in 2\mathbb{Z}\ {\rm for}\ \forall i) \\
2s{\rm det}N' & ({\rm otherwise})
\end{array}
\right., \\
\ [ \rho_{T^{2g}}(\widetilde{T}_{X}) \rho_{T^{2g}}(\widetilde{T}_{Y}) ]_{JK} = [ \rho_{T^{2g}}(\widetilde{T}_{Y}) \rho_{T^{2g}}(\widetilde{T}_{X}) ]_{JK},
\end{array}
\label{Ttildewaverel}
\end{align}
where $T_{X}$ and $T_{Y}$ denote $T$ transformations with $B=B_{X}$ and $B=B_{Y}$, respectively.
Thus, the wave functions on magnetized $T^{2g}$ in Eq.~(\ref{wavT2g}) behave as the Siegel modular forms of weight $1/2$ and $\widetilde{N}_{g}(H,h)$, and then they transform non-trivially under $\widetilde{N}_{g,h}(H)=\widetilde{N}_{g}(H)/\widetilde{N}_{g}(H,h)$ modular flavor transformation, where $h$ denotes the least common multiple of all orders of $h_{X}$.
In particular, we denote $\widetilde{N}_{g}^{(g-1-a)}(H) \equiv \widetilde{\Gamma}_{g}$, $\widetilde{N}_{g}^{(g-1-a)}(H,h) \equiv \widetilde{\Gamma}_{g}(h)$, and then $\widetilde{N}_{g,h}^{(g-1-a)} \equiv \widetilde{\Gamma}_{g,h}$, with $h=2n$.\footnote{For $g=1$, it has been studied in Ref.~\cite{Kikuchi:2020frp}.}

Furthermore, when we consider $T^{2g}$ orbifold constructed by identifying a stabilizer $A_X \in \mathbb{Z}_2$ transformation (e.g. discussed in Ref.~\cite{Kikuchi:2022lfv}\footnote{See also Refs.~\cite{Kikuchi:2022psj,Kikuchi:2023awm}.}), the wave functions on the magnetized $T^{2g}/\mathbb{Z}_2$, $\psi^{J,N}_{T^{2g}/\mathbb{Z}_2^m}(z,\Omega)$, which satisfy the boundary condition for $A_X \in \mathbb{Z}_2$ transformation,
\begin{align}
\psi^{J,N}_{T^{2g}/\mathbb{Z}_2^m}(X z,\Omega) = (-1)^m \psi^{J,N}_{T^{2g}/\mathbb{Z}_2^m}(z,\Omega), \label{BCPsiXz}
\end{align}
as well as the boundary conditions in Eqs.~(\ref{BCPsiz1}) and (\ref{BCPsiztau}),
can be expanded by wave functions on the magnetized $T^{2g}$ as
\begin{align}
\psi^{J,N}_{T^{2g}/\mathbb{Z}_2^m}(z,\Omega) = {\cal N}_{T^{2g}/\mathbb{Z}_2} \left( \psi^{J,N}_{T^{2g}}(z,\Omega) + (-1)^m \psi^{{^t}X J,N}_{T^{2g}}(z,\Omega) \right),
\end{align}
and then they transform under the $A_{X}$ transformation as
\begin{align}
\begin{array}{l}
\psi^{J,N}_{T^{2g}}(\widetilde{A}_{X}^{\pm}(z,\Omega)) = \widetilde{J}_{1/2}(\widetilde{A}_{X}^{\pm}, \Omega) \rho_{T^{2g}/\mathbb{Z}_2^{m}}(\widetilde{A}_{X}^{\pm})_{JK} \psi^{K,N}_{T^{2g}/\mathbb{Z}_2^m}(z,\Omega) = (-1)^m \psi^{J,N}_{T^{2g}/\mathbb{Z}_2^m}(z,\Omega), \\
\widetilde{A}_{X}^{\pm} \equiv [A_{X},\pm 1] : \ \widetilde{J}_{1/2}(\widetilde{A}_{X}^{\pm},\Omega) = \pm ({\rm det}X)^{1/2}, \ \rho_{T^{2g}/\mathbb{Z}_2^{m}}(\widetilde{A}_{X}^{\pm})_{JK} = \pm (-1)^m ({\rm det}X)^{-1/2} \delta_{J,K},
\end{array}
\label{AtildewaveZ2rel}
\end{align}
where $m$ and ${\cal N}_{T^{2g}/\mathbb{Z}_2} $denote the eigenvalue of $A_X \in \mathbb{Z}_2$ transformation and the normalization factor such that Eq.~(\ref{Normalization}) is also satisfied, respectively. 
Similarly, we can find that the behavior of wave functions on the magnetized $T^{2g}/\mathbb{Z}_2$ under the modular transformation is the same as that of wave functions on the magnetized $T^{2g}$.
The difference is just the basis of the representation.
In particular, in the orbifold eigenbasis, the representation can be block diagonalized by orbifold eigenvalues.
In other words, once the orbifold eigenvalue is fixed, 
we can obtain smaller representation in the orbifold case
than the $T^{2g}$ case, in general.
Hence, we basically consider the orbifold case, hereafter.

Now, let us see concrete modular flavor symmetry of lower dimensional (two, three, and four dimensional) wave functions on magnetized $T^{2g}$ orbifolds.
In particular, we mainly discuss $g=2$ cases, 
For $g=1$ cases, we have studied $g=1$ cases in Refs.~\cite{Kikuchi:2020frp,Kikuchi:2021ogn}.
On the other hand, while a lot of $g=3$ cases can be studied in ways similar to  $g=2$ cases, the generation numbers become larger in general, and then the modular flavor groups become larger and the analysis becomes more complicated.
Hereafter, we often use the following notation,
\begin{align}
\ket{J_1,...,J_g} \equiv \psi^{J,N}_{T^{2g}}(z,\Omega) \quad {\rm with}\quad J={^t}(J_1,...,J_g).
\end{align}
At first, let us see the class ($g$-1-a) with $n \in 2\mathbb{Z}$.
Wave functions on magnetized $T^{2g}$ as well as $T^{2g}/\mathbb{Z}_2^{t}$ transform non-trivially under $\widetilde{\Gamma}_{g,2n}$ modular flavor transformation.
For example, $n=2$ is the minimal example.
For $g=1$ case, in Ref.~\cite{Kikuchi:2021ogn},\footnote{See also Ref.~\cite{Uchida:2023yaj}.} it was found that
two-dimensional $\mathbb{Z}_2^t$ twisted even $(m^t=0)$ modes on magnetized $T^2/\mathbb{Z}_2^t$ transform non-trivially under $\widetilde{\Gamma}_{1,4} \simeq T' \rtimes Z_4 \simeq \widetilde{S}_4$~\cite{Liu:2020msy}.
Let us see the $g=2$ case,~i.e.,
\begin{align}
\begin{array}{l}
N=
\begin{pmatrix}
2 & 0 \\
0 & 2
\end{pmatrix}
=2I_{2}, \\
\Omega=
\begin{pmatrix}
\tau_{11} & \tau_{12} \\
\tau_{12} & \tau_{22}
\end{pmatrix}
=\sum_{i,j=1,2,3} \tau_{ij} B_{ij}.
\end{array}
\label{2002}
\end{align}
In this case, the following four-dimensional $\mathbb{Z}_2^t$ twisted even $(m^t=0)$ modes\footnote{The four-dimensional modes can be modular forms at $z=0$.} on magnetized $T^2/\mathbb{Z}_2^t$,
\begin{align}
\ket{J_1,J_2} =
\begin{pmatrix}
\ket{0,0} \\ \ket{1,0} \\ \ket{0,1} \\ \ket{1,1}
\end{pmatrix},
\label{class21afour}
\end{align}
transform under $S$ and $T_{ab}$ transformations as
\begin{align}
\rho_{T^{4}/\mathbb{Z}_2^{0}}(\widetilde{S}) = \frac{i}{2}
\begin{pmatrix}
1 & 1 & 1 & 1 \\
1 & -1 & 1 & -1 \\
1 & 1 & -1 & -1 \\
1 & -1 & -1 & 1
\end{pmatrix}, &\quad
\rho_{T^{4}/\mathbb{Z}_2^{0}}(\widetilde{T}_{12}) = 
\begin{pmatrix}
1 & & &  \\
 & 1 & & \\
 & & 1 & \\
 & & & -1
\end{pmatrix}, \notag \\
\rho_{T^{4}/\mathbb{Z}_2^{0}}(\widetilde{T}_{11}) = 
\begin{pmatrix}
1 & & &  \\
 & i & & \\
 & & 1 & \\
 & & & i
\end{pmatrix}, &\quad
\rho_{T^{4}/\mathbb{Z}_2^{0}}(\widetilde{T}_{22}) = 
\begin{pmatrix}
1 & & &  \\
 & 1 & & \\
 & & i & \\
 & & & i
\end{pmatrix}, \label{20024genrho}
\end{align}
which satisfy the following relations,
\begin{align}
\begin{array}{l}
\ \rho_{T^{4}/\mathbb{Z}_2^{0}}(\widetilde{S})^2 = -I, \\
\ [\rho_{T^{4}/\mathbb{Z}_2^{0}}(\widetilde{S}) \rho_{T^{4}/\mathbb{Z}_2^{0}}(\widetilde{T}_{12})]^6 = - I, \\
\ [\rho_{T^{4}/\mathbb{Z}_2^{0}}(\widetilde{S}) \rho_{T^{4}/\mathbb{Z}_2^{0}}(\widetilde{T}_{11})]^6 = [\rho_{T^{4}/\mathbb{Z}_2^{0}}(\widetilde{S}) \rho_{T^{4}/\mathbb{Z}_2^{0}}(\widetilde{T}_{22})]^6 = -i I, \\
\ \rho_{T^{4}/\mathbb{Z}_2^{0}}(\widetilde{T}_{12})^2 = I, \quad \rho_{T^{4}/\mathbb{Z}_2^{0}}(\widetilde{T}_{11})^4 = \rho_{T^{4}/\mathbb{Z}_2^{0}}(\widetilde{T}_{22})^4 = I, \\
\ \rho_{T^{4}/\mathbb{Z}_2^{0}}(\widetilde{T}_{ab}) \rho_{T^{4}/\mathbb{Z}_2^{0}}(\widetilde{T}_{cd}) = \rho_{T^{4}/\mathbb{Z}_2^{0}}(\widetilde{T}_{cd}) \rho_{T^{4}/\mathbb{Z}_2^{0}}(\widetilde{T}_{ab}) \quad (a,b,c,d \in \{1,2\}).
\end{array}
\end{align}
Then, they transform non-trivially under $\widetilde{\Gamma}_{2,4}$ modular flavor transformation.~\footnote{The order is too large to specify the concrete modular flavor group.}
We note that we can similarly consider the $g=3$ case; eight-dimensional $\mathbb{Z}_2^t$ twisted even $(m^t=0)$ modes on magnetized $T^6/\mathbb{Z}_2$ with $N=2I_{3}$ transform non-trivially under $\widetilde{\Gamma}_{3,4}$ modular flavor transformation.
Moreover, by combining the classes (1-1-a) and (2-1-a), we can similarly consider the classes (2-1-b) and (3-1-b).
In the above example, if $\tau_{12}$ is restricted to $\tau_{12}=0$, it corresponds to the specific case of the class (2-1-b).
Hence, the four-dimensional $\mathbb{Z}_2^{t_k}$ twisted even $(m^{t_k}=0, \ k=1,2)$ modes on magnetized  $T^2/\mathbb{Z}_2^{t_1} \times T^2/\mathbb{Z}_2^{t_2}$ in Eq.~(\ref{class21afour}) transform non-trivially under $\widetilde{\Gamma}_{1,4} \times \widetilde{\Gamma}_{1,4}=\widetilde{S}_4 \times \widetilde{S}_4$ modular flavor transformation.
Furthermore, if there is a constraint between $\tau_{11}$ and $\tau_{22}$ such that $\tau_{11}=\tau_{22}=\tau$ in addition to $\tau_{12}=0$,~i.e. $\Omega=\tau I$, the three-dimensional $\mathbb{Z}_2^{t}$ twisted even $(m^{t}=0)$ and $\mathbb{Z}_2^{p}$ permutation even $(m^{p}=0)$ modes\footnote{The three-dimensional modes can be modular forms at $z=0$.} on magnetized $T^2/(\mathbb{Z}_2^{t} \times \mathbb{Z}_2^{m})$,
\begin{align}
\ket{J_1,J_2} =
\begin{pmatrix}
\ket{0,0} \\ \frac{1}{\sqrt{2}} (\ket{1,0} + \ket{0,1} ) \\ \ket{1,1}
\end{pmatrix},
\label{class21athree}
\end{align}
transform non-trivially under $\Gamma'_{1,4} \simeq S'_4 \simeq \Delta'(24) \simeq [(Z_2 \times Z'_2) \rtimes Z_3] \rtimes Z_4$ modular flavor transformation with weight $1$~\cite{Kikuchi:2020nxn,Kikuchi:2021ogn}, where we can write the generators of $Z_2$, $Z'_2$, $Z_3$, and $Z_4$ as
\begin{align}
a=T^2ST^2ST^2, \ a'=ST^2S^{-1}T^{-2}, \ b=TS^3T^2, \ c=ST^2ST^3, \label{S'4gen}
\end{align}
respectively, and we can indeed check the relations among them;
\begin{align}
\begin{array}{l}
a^2=a'^2=b^3=c^4=1,\ (c=-1) \\
aa'=a'a, \ bab^{-1}=a^{-1}a'^{-1}, \ ba'b^{-1}=a, \ cac^{-1}=a'^{-1}, \ ca'c^{-1}=a^{-1}, \ cbc^{-1}=b^{-1}. \label{S'4alg}
\end{array}
\end{align}
On the other hand, if the constraints are $\tau_{11}=\tau_{22}=\tau$ and $\tau_{12} \neq 0$, it corresponds to the specific case of the class (2-2-a).
Hence, the three-dimensional modes on magnetized $T^2/(\mathbb{Z}_2^{t} \times \mathbb{Z}_2^{m})$ in Eq.~(\ref{class21athree}) transform under $S$, $T_{I_2}=T_{11}T_{22}$ and $T_{12}$ transformations as
\begin{align}
&\rho_{T^{4}/(\mathbb{Z}_2^{0_t} \times \mathbb{Z}_2^{0_p})}(\widetilde{S}) = \frac{i}{2}
\begin{pmatrix}
1 & \sqrt{2} & 1 \\
\sqrt{2} & 0 & -\sqrt{2} \\
1 & -\sqrt{2} & 1
\end{pmatrix}, \notag \\
&\rho_{T^{4}/(\mathbb{Z}_2^{0_t} \times \mathbb{Z}_2^{0_p})}(\widetilde{T}_{I_2}) =
\begin{pmatrix}
1 & & \\
 & i & \\
 & & -1
\end{pmatrix}, \quad
\rho_{T^{4}/(\mathbb{Z}_2^{0_t} \times \mathbb{Z}_2^{0_p})}(\widetilde{T}_{12}) =
\begin{pmatrix}
1 & & \\
 & 1 & \\
 & & -1
\end{pmatrix},
\label{2002rho3}
\end{align}
which satisfy the following relations,
\begin{align}
\begin{array}{l}
\ \rho_{T^{4}/(\mathbb{Z}_2^{0_t} \times \mathbb{Z}_2^{0_p})}(\widetilde{S})^2 = -I, \\
\ [ \rho_{T^{4}/(\mathbb{Z}_2^{0_t} \times \mathbb{Z}_2^{0_p})}(\widetilde{S}) \rho_{T^{4}/(\mathbb{Z}_2^{0_t} \times \mathbb{Z}_2^{0_p})}(\widetilde{T}_{I_2}) ]^3 = I, \\
\ [ \rho_{T^{4}/(\mathbb{Z}_2^{0_t} \times \mathbb{Z}_2^{0_p})}(\widetilde{S}) \rho_{T^{4}/(\mathbb{Z}_2^{0_t} \times \mathbb{Z}_2^{0_p})}(\widetilde{T}_{12}) ]^3 = -i I, \\
\ \rho_{T^{4}/(\mathbb{Z}_2^{0_t} \times \mathbb{Z}_2^{0_p})}(\widetilde{T}_{12}) ^2 = I, \quad \rho_{T^{4}/(\mathbb{Z}_2^{0_t} \times \mathbb{Z}_2^{0_p})}(\widetilde{T}_{I_2})^4 = I, \\
\ \rho_{T^{4}/(\mathbb{Z}_2^{0_t} \times \mathbb{Z}_2^{0_p})}(\widetilde{T}_{I_2}) \rho_{T^{4}/(\mathbb{Z}_2^{0_t} \times \mathbb{Z}_2^{0_p})}(\widetilde{T}_{12}) =\rho_{T^{4}/(\mathbb{Z}_2^{0_t} \times \mathbb{Z}_2^{0_p})}(\widetilde{T}_{12}) \rho_{T^{4}/(\mathbb{Z}_2^{0_t} \times \mathbb{Z}_2^{0_p})}(\widetilde{T}_{I_2}).
\end{array}
\end{align}
Then, they transform under $\widetilde{N}_{2,4}^{(2-2-a)}(H)$ modular flavor transformation.
Actually, when we define
\begin{align}
\begin{array}{l}
\tilde{a}=T_{I_2}T_{12}ST_{12}T_{I_2}S^{-1}T_{I_2}T_{12} (ST_{12})^3, \ \tilde{a}'= ST_{I_2}T_{12}S^{-1}T_{12}^{-1}T_{I_2}^{-1}, \\ b=T_{I_2}S^3T_{I_2}^2, \ c'=ST_{I_2}^2ST_{I_2}^3 (ST_{12})^3, \\
d = (ST_{12})^3,
\end{array}
\label{Delta96Z4gen}
\end{align}
we can find that they satisfy the following relations,
\begin{align}
\begin{array}{l}
\tilde{a}^4=\tilde{a}'^4=b^3=c'^2=d^4=1, \\
\tilde{a}\tilde{a}'=\tilde{a}'\tilde{a}, \ b\tilde{a}b^{-1}=\tilde{a}^{-1}\tilde{a}'^{-1}, \ b\tilde{a}'b^{-1}=\tilde{a}, \ c'\tilde{a}c'^{-1}=\tilde{a}'^{-1}, \ c'\tilde{a}'c'^{-1}=\tilde{a}^{-1}, \ c'bc'^{-1}=b^{-1}, \\
dx=xd\ (x=\tilde{a}, \tilde{a}', b, c'),
\end{array}
\label{Delta96Z4alg}
\end{align}
that is, the three-dimensional modes in Eq.~(\ref{class21athree}) transform non-trivially under $\Delta(96) \times Z_4$ modular flavor transformation.
Notice that we can find the following relation between $\tilde{a}^{(\prime)}$ in Eq.~(\ref{Delta96Z4gen}) and $a^{(\prime)}$ in Eq.~(\ref{S'4gen}),
\begin{align}
(\tilde{a}^{(\prime)})^2 = a^{(\prime)}. \label{atildea}
\end{align}
We can check it in Appendix~\ref{ap:proof}.
Therefore, we can obtain two patterns of breaking chains,~i.e.,
\begin{align}
\begin{array}{rlcrl}
 & {_{\tau_{12}=0}} \ & \widetilde{\Gamma}_{2,4} & \ {_{\tau_{11}=\tau_{22}}} & \\
 & \quad \swarrow & & \searrow \quad & \\
\widetilde{S}_4 \times \widetilde{S}_4 & & & & \Delta(96) \times Z_4 \\
 & \quad \searrow & & \swarrow \quad & \\
 & {^{\tau_{11}=\tau_{22}}} \ & S'_4 & \ {^{\tau_{12}=0}} & 
\end{array}.
\label{chain}
\end{align}

As another example of the class (2-2-a), let us consider the case that
\begin{align}
\begin{array}{l}
N =
\begin{pmatrix}
4 & -2 \\
-2 & 4
\end{pmatrix}
= 2
\begin{pmatrix}
2 & -1 \\
-1 & 2
\end{pmatrix}
= 2 N', \quad
{\rm det}N = 4{\rm det}N' = 12, 
\quad
\tilde{N} = 2\tilde{N} = 2
\begin{pmatrix}
2 & 1 \\
1 & 2
\end{pmatrix}, 
\\
\Omega=
\begin{pmatrix}
\tau & \tau_{12} \\
\tau_{12} & \tau
\end{pmatrix}
=\tau B_{I_2} + \tau_{12} B_{12}.
\end{array}
\label{4224}
\end{align}
In this case, the following three-dimensional $\mathbb{Z}_2^t$ twisted odd $(m^t=1)$ and $\mathbb{Z}_2^p$ permutation odd $(m^p=1)$ modes,
\begin{align}
\ket{J_1,J_2} =
\begin{pmatrix}
\frac{1}{\sqrt{2}} (\ket{-1,3}-\ket{3,-1}) \\  
\frac{1}{2} (\ket{1,0}-\ket{0,1}+\ket{2,1}-\ket{1,2}) \\
\frac{1}{\sqrt{2}} (\ket{2,0}-\ket{0,2}) \\ 
\end{pmatrix},
\label{42243gen}
\end{align}
transform under $S$, $T_{I_2}$, and $T_{12}$ transformations as
\begin{align}
&\rho_{T^{4}/(\mathbb{Z}_2^{1_t} \times \mathbb{Z}_2^{1_p})}(\widetilde{S}) = -\frac{1}{2}
\begin{pmatrix}
1 & \sqrt{2} & 1 \\
\sqrt{2} & 0 & -\sqrt{2} \\
1 & -\sqrt{2} & 1
\end{pmatrix}, \notag \\
&\rho_{T^{4}/(\mathbb{Z}_2^{1_t} \times \mathbb{Z}_2^{1_p})}(\widetilde{T}_{I_2}) =
\begin{pmatrix}
e^{\pi i/3} & & \\
 & e^{\pi i/3} & \\
 & & e^{4\pi i/3}
\end{pmatrix}
= e^{\pi i/3}
\begin{pmatrix}
1 & & \\
 & 1 & \\
 & & -1
\end{pmatrix}, \\
&\rho_{T^{4}/(\mathbb{Z}_2^{1_t} \times \mathbb{Z}_2^{1_p})}(\widetilde{T}_{12}) =
\begin{pmatrix}
e^{-\pi i/3} & & \\
 & e^{\pi i/6} & \\
 & & e^{2\pi i/3}
\end{pmatrix}
= e^{-\pi i/3}
\begin{pmatrix}
1 & & \\
 & i & \\
 & & -1
\end{pmatrix}, \notag
\end{align}
which satisfy the following relations,
\begin{align}
\begin{array}{l}
\ \rho_{T^{4}/(\mathbb{Z}_2^{1_t} \times \mathbb{Z}_2^{1_p})}(\widetilde{S})^2 = I, \\
\ [ \rho_{T^{4}/(\mathbb{Z}_2^{1_t} \times \mathbb{Z}_2^{1_p})}(\widetilde{S}) \rho_{T^{4}/(\mathbb{Z}_2^{1_t} \times \mathbb{Z}_2^{1_p})}(\widetilde{T}_{I_2}) ]^3 = I, \\
\ [ \rho_{T^{4}/(\mathbb{Z}_2^{1_t} \times \mathbb{Z}_2^{1_p})}(\widetilde{S}) \rho_{T^{4}/(\mathbb{Z}_2^{1_t} \times \mathbb{Z}_2^{1_p})}(\widetilde{T}_{12}) ]^3 = i I, \\
\ \rho_{T^{4}/(\mathbb{Z}_2^{1_t} \times \mathbb{Z}_2^{1_p})}(\widetilde{T}_{I_2})^2 = 
\rho_{T^{4}/(\mathbb{Z}_2^{1_t} \times \mathbb{Z}_2^{1_p})}(\widetilde{T}_{12})^4 = e^{2\pi i/3} I, \\
\ \rho_{T^{4}/(\mathbb{Z}_2^{1_t} \times \mathbb{Z}_2^{1_p})}(\widetilde{T}_{I_2}) \rho_{T^{4}/(\mathbb{Z}_2^{1_t} \times \mathbb{Z}_2^{1_p})}(\widetilde{T}_{12}) =\rho_{T^{4}/(\mathbb{Z}_2^{1_t} \times \mathbb{Z}_2^{1_p})}(\widetilde{T}_{12}) \rho_{T^{4}/(\mathbb{Z}_2^{1_t} \times \mathbb{Z}_2^{1_p})}(\widetilde{T}_{I_2}).
\end{array}
\end{align}
Then, they transform non-trivially under $\widetilde{N}_{2,12}^{(2-2-a)}(H)$ modular flavor transformation.
We note that wave functions in the class (2-2-a) generally transform non-trivially under $\widetilde{N}_{2,2s{\rm det}N'}^{(2-2-a)}(H)$ modular flavor transformation.
Actually, similar to Eq.~(\ref{2002rho3}), we can find that the three-dimensional modes in Eq.~(\ref{42243gen}), transform non-trivially under $\Delta(96) \times Z_4 \times Z_3$ modular flavor transformation, where the generators of $\Delta(96) \times Z_4$ are in Eq.~(\ref{Delta96Z4gen}) replacing $S$, $T_{I_2}$, and $T_{12}$ with the following $s$, $t$, and $t_{12}$,
\begin{align}
s=S (ST_{12})^{-3}, \ t=T_{12} T_{I_2}^{-2} (ST_{12})^{-6}, \ t_{12} = T_{I_2} T_{12}^{4} (ST_{12})^{6},
\label{4224stt12}
\end{align}
respectively, while the $Z_3$ generator is $e=T_{I_2}^2=T_{12}^4$, satisfying
\begin{align}
e^3=1, \ ex=xe\ (x=\tilde{a}, \tilde{a}', b, c', d).
\end{align}
Here, if $\tau_{12}$ is restricted to $\tau_{12}=0$,~i.e. $\Omega=\tau I$, the three-dimensional modes in Eq.~(\ref{42243gen}) transform non-trivially under $S_3 \times Z_3 \simeq (Z'_3 \rtimes Z_2) \times Z_3$ modular flavor transformation with weight $1$, where the generators $Z'_3$, $Z_2$, and $Z_3$ are written as
\begin{align}
p=ST_{I_2}, \ q=T_{I_2}^3, \ e=T_{I_2}^2,
\end{align}
satisfying
\begin{align}
p^3=q^2=e^3=1, \ q^{-1}pq=p^2, \ ex=xe\ (x=p,q).
\end{align}
Therefore, we can obtain the following breaking pattern,
\begin{align}
\Delta(96) \times Z_4 \times Z_3 \xrightarrow{\tau_{12}=0} S_3 \times Z_3. \label{Delta96Z4Z3S3Z3}
\end{align}
So far, we have seen the modular flavor symmetry of  three-dimensional $\mathbb{Z}_2^t$ twisted odd $(m^t=1)$ and $\mathbb{Z}_2^p$ permutation odd $(m^p=1)$ modes in Eq.~(\ref{42243gen}).
On the other hand, the following two-dimensional $\mathbb{Z}_2$ twisted even $(m^t=0)$ and $\mathbb{Z}_2$ permutation odd $(m^p=1)$ modes,
\begin{align}
\ket{J_1,J_2} =
\begin{pmatrix}
\frac{1}{2} (\ket{1,0}-\ket{0,1}-\ket{2,1}+\ket{1,2}) \\
\frac{1}{\sqrt{2}} (\ket{2,-1}-\ket{-1,2})
\end{pmatrix},
\label{42242gen}
\end{align}
transform under $S$, $T_{I_2}$, and $T_{12}$ transformations as
\begin{align}
&\rho_{T^{4}/(\mathbb{Z}_2^{0_t} \times \mathbb{Z}_2^{1_p})}(\widetilde{S}) = -\frac{i}{\sqrt{3}}
\begin{pmatrix}
1 & \sqrt{2} \\
\sqrt{2} & -1
\end{pmatrix}, \notag \\
&\rho_{T^{4}/(\mathbb{Z}_2^{0_t} \times \mathbb{Z}_2^{1_p})}(\widetilde{T}_{I_2}) =
\begin{pmatrix}
e^{\pi i/3} & \\
 & -1
\end{pmatrix}, \quad 
\rho_{T^{4}/(\mathbb{Z}_2^{0_t} \times \mathbb{Z}_2^{1_p})}(\widetilde{T}_{12}) =
\begin{pmatrix}
e^{\pi i/6} & \\
 & -i  \\
\end{pmatrix},
\end{align}
which satisfy the following relations,
\begin{align}
\begin{array}{l}
\ \rho_{T^{4}/(\mathbb{Z}_2^{0_t} \times \mathbb{Z}_2^{1_p})}(\widetilde{S})^2 = -I, \\
\ [ \rho_{T^{4}/(\mathbb{Z}_2^{0_t} \times \mathbb{Z}_2^{1_p})}(\widetilde{S}) \rho_{T^{4}/(\mathbb{Z}_2^{0_t} \times \mathbb{Z}_2^{1_p})}(\widetilde{T}_{I_2}) ]^3 = I, \\
\ [ \rho_{T^{4}/(\mathbb{Z}_2^{1_t} \times \mathbb{Z}_2^{0_p})}(\widetilde{S}) \rho_{T^{4}/(\mathbb{Z}_2^{0_t} \times \mathbb{Z}_2^{1_p})}(\widetilde{T}_{12}) ]^3 = i I, \\
\ \rho_{T^{4}/(\mathbb{Z}_2^{0_t} \times \mathbb{Z}_2^{1_p})}(\widetilde{T}_{I_2}) = \rho_{T^{4}/(\mathbb{Z}_2^{0_t} \times \mathbb{Z}_2^{1_p})}(\widetilde{T}_{12})^{2}, \\
\ \rho_{T^{4}/(\mathbb{Z}_2^{0_t} \times \mathbb{Z}_2^{1_p})}(\widetilde{T}_{12})^{3} = i I, \\
\ \rho_{T^{4}/(\mathbb{Z}_2^{0_t} \times \mathbb{Z}_2^{1_p})}(\widetilde{T}_{I_2})^3 = \rho_{T^{4}/(\mathbb{Z}_2^{0_t} \times \mathbb{Z}_2^{1_p})}(\widetilde{T}_{12})^{6} = -I, \\
\ \rho_{T^{4}/(\mathbb{Z}_2^{0_t} \times \mathbb{Z}_2^{1_p})}(\widetilde{T}_{I_2}) \rho_{T^{4}/(\mathbb{Z}_2^{0_t} \times \mathbb{Z}_2^{1_p})}(\widetilde{T}_{12}) =\rho_{T^{4}/(\mathbb{Z}_2^{0_t} \times \mathbb{Z}_2^{1_p})}(\widetilde{T}_{12}) \rho_{T^{4}/(\mathbb{Z}_2^{0_t} \times \mathbb{Z}_2^{1_p})}(\widetilde{T}_{I_2}).
\end{array}
\end{align}
Then, they transform non-trivially under $\widetilde{N}_{2,12}^{(2-2-a)}(H)$ modular flavor transformation.
Actually, when we define
\begin{align}
s = S^{-1}, \ t = S^2T_{I_2}, \ 
c = ST_{I_2}T_{12},
\label{T'Z2gen}
\end{align}
we can find that they satisfy the following relations,
\begin{align}
s^2=-1, \ s^4=t^3=(st)^3=c^2=1, \ c^{-1} x c = x'\ (x, x' \in T').
\label{T'Z2alg}
\end{align}
Thus, the two-dimensional modes in Eq.~(\ref{42242gen}) transform non-trivially under $T' \rtimes Z_2$ modular flavor transformation.
Here, if $\tau_{12}$ is restricted to $\tau_{12}=0$,~i.e. $\Omega=\tau I$, the two-dimensional modes in Eq.~(\ref{42242gen}) transform non-trivially under $T'$ modular flavor transformation with weight $1$, where the generators are $s$ and $t$ in Eq.~(\ref{T'Z2gen}).
Therefore, we can obtain the following breaking pattern,
\begin{align}
T' \rtimes Z_2 \xrightarrow{\tau_{12}=0} T'.
\end{align}

Similarly, let us see the following example of the class (2-2-b) with $\Delta N_{12}=N_{12}$ case,
\begin{align}
\begin{array}{l}
N =
\begin{pmatrix}
2 & -2 \\
-2 & 4
\end{pmatrix}
= 2
\begin{pmatrix}
1 & -1 \\
-1 & 2
\end{pmatrix}
= 2 N', \quad
{\rm det}N = 4{\rm det}N' = 4, \quad
\tilde{N} = 2\tilde{N} = 2
\begin{pmatrix}
1 & -1 \\
-1 & 2
\end{pmatrix}, 
\\
\Omega=
\begin{pmatrix}
\tau+\tau_{N} & \tau_{N} \\
\tau_{N} & \tau
\end{pmatrix}
=\tau B_{I_2} + \tau_{N} B_{N_2}.
\end{array}
\label{2224}
\end{align}
In this case, the following four-dimensional $\mathbb{Z}_2^t$ twisted even $(m^t=0)$ modes\footnote{The four-dimensional modes can be modular forms at $z=0$.},
\begin{align}
\ket{J_1,J_2} =
\begin{pmatrix}
\ket{0,0} \\
\ket{0,1} \\
\ket{1,-1} \\
\ket{-1,2}
\end{pmatrix},
\label{22244gen}
\end{align}
transform under $S$, $T_{I_2}$, and $T_{N_2}$ transformations as
\begin{align}
&\rho_{T^{4}/\mathbb{Z}_2^{0}}(\widetilde{S}) = \frac{i}{2}
\begin{pmatrix}
1 & 1 & 1 & 1 \\
1 & -1 & 1 & -1 \\
1 & 1 & -1 & -1 \\
1 & -1 & -1 & 1
\end{pmatrix}, \notag \\
&\rho_{T^{4}/\mathbb{Z}_2^{0}}(\widetilde{T}_{I_2}) = 
\begin{pmatrix}
1 & & &  \\
 & i & & \\
 & & i & \\
 & & & -1
\end{pmatrix}, \label{22244genrho} \quad
\rho_{T^{4}/\mathbb{Z}_2^{0}}(\widetilde{T}_{N_2}) = 
\begin{pmatrix}
1 & & &  \\
 & i & & \\
 & & 1 & \\
 & & & -i
\end{pmatrix},
\end{align}
which satisfy the following relations,
\begin{align}
\begin{array}{l}
\ \rho_{T^{4}/\mathbb{Z}_2^{0}}(\widetilde{S})^2 = -I, \\
\ [\rho_{T^{4}/\mathbb{Z}_2^{0}}(\widetilde{S}) \rho_{T^{4}/\mathbb{Z}_2^{0}}(\widetilde{T}_{I_2})]^3 = I, \\
\ [\rho_{T^{4}/\mathbb{Z}_2^{0}}(\widetilde{S}) \rho_{T^{4}/\mathbb{Z}_2^{0}}(\widetilde{T}_{N_2})]^5 = - I, \\
\ \rho_{T^{4}/\mathbb{Z}_2^{0}}(\widetilde{T}_{I_2})^4 = I, \quad \rho_{T^{4}/\mathbb{Z}_2^{0}}(\widetilde{T}_{N_2})^4 = I, \\
\ \rho_{T^{4}/\mathbb{Z}_2^{0}}(\widetilde{T}_{I_2}) \rho_{T^{4}/\mathbb{Z}_2^{0}}(\widetilde{T}_{N_2}) = \rho_{T^{4}/\mathbb{Z}_2^{0}}(\widetilde{T}_{N_2}) \rho_{T^{4}/\mathbb{Z}_2^{0}}(\widetilde{T}_{I_2}).
\end{array}
\end{align}
Then, they transform non-trivially under $\widetilde{N}_{2,4}^{(2-2-b)}(H)$ modular flavor transformation.
We note that wave functions in the class (2-2-b) generally transform non-trivially under $\widetilde{N}_{2,2s{\rm det}N'}^{(2-2-b)}(H)$ modular flavor transformation.
Here, the above $S$, $T_{I_2}=T_{11}T_{22}$, and $T_{N_2}=T_{11}T_{12}$ transformations in Eq.~(\ref{22244genrho}) correspond to the $S$, $T_{11}T_{22}$, and $T_{11}T_{12}$ transformations in Eq.~(\ref{20024genrho}), respectively.
Actually,
we can find that
the four-dimensional modes in Eq.~(\ref{22244gen}) transform non-trivially under $[(Z_4 \times Z_2 \times Z_2) \rtimes (Z_2 \times Z_2)] \rtimes A_5$ modular flavor transformation.
If $\tau_{N}$ is restricted to $\tau_{N}=0$,~i.e. $\Omega=\tau I$, the four-dimensional modes in Eq.~(\ref{22244gen}) transform non-trivially under $\widetilde{\Gamma}'_{1,4} \simeq S'_4$ modular flavor transformation with weight $1$.
Therefore, we can obtain the following breaking pattern,
\begin{align}
[(Z_4 \times Z_2 \times Z_2) \rtimes (Z_2 \times Z_2)] \rtimes A_5 \xrightarrow{\tau_{N}=0} S'_4.
\end{align}

Now, we can similarly consider the case of the class (3-2-a).
As a specific case, let us see the following example of the class (3-2-b) with $N_{11}=N_{22}=n=4$ and $N_{33}=n + N_{12}=4-2=2$ case,~i.e.,
\begin{align}
\begin{array}{l}
N =
\begin{pmatrix}
4 & -2 & 0 \\
-2 & 4 & 0 \\
0 & 0 & 2
\end{pmatrix}
= 2
\begin{pmatrix}
2 & -1 & 0 \\
-1 & 2 & 0 \\
0 & 0 & 1
\end{pmatrix}
= 2 N', 
\\
\Omega=
\begin{pmatrix}
\tau & \tau_{12} & \tau' \\
\tau_{12} & \tau & \tau' \\
\tau' & \tau' & \tau_{33}
\end{pmatrix}
=\tau B_{I_2} + \tau_{12} B_{12} + \tau_{33} B_{33} + \tau' B'_{+}.
\end{array}
\label{42242}
\end{align}
In this case, the following four-dimensional $\mathbb{Z}_2^t$ twisted even $(m^t=0)$ and $\mathbb{Z}_2^p$ permutation odd ($m^p=1$) modes,
\begin{align}
&\rho_{T^{6}/(\mathbb{Z}_2^{0_t} \times \mathbb{Z}_2^{1_p})}(\widetilde{S}) = \frac{e^{3\pi i/4}}{\sqrt{6}}
\begin{pmatrix}
1 & \sqrt{2} & 1 & \sqrt{1} \\
\sqrt{2} & -1 & \sqrt{2} & -1 \\
1 & \sqrt{2} & -1 & -\sqrt{2} \\
\sqrt{2} & -1 & -\sqrt{2} & 1
\end{pmatrix}, \notag \\
&\rho_{T^{6}/(\mathbb{Z}_2^{0_t} \times \mathbb{Z}_2^{1_p})}(\widetilde{T}_{I_2}) =
\begin{pmatrix}
e^{\pi i/3} & & & \\
 & -1 & & \\
 & & e^{\pi i/3} & \\
 & & & -1
\end{pmatrix}, \quad 
\rho_{T^{4}/(\mathbb{Z}_2^{0_t} \times \mathbb{Z}_2^{1_p})}(\widetilde{T}_{12}) =
\begin{pmatrix}
e^{\pi i/6} & & & \\
 & -i & & \\
 & & e^{\pi i/6} & \\
 & & & -i
\end{pmatrix}, \notag \\
&\rho_{T^{6}/(\mathbb{Z}_2^{0_t} \times \mathbb{Z}_2^{1_p})}(\widetilde{T}_{33}) =
\begin{pmatrix}
1 & & & \\
 & 1 & & \\
 & & i & \\
 & & & i
\end{pmatrix}, \quad 
\rho_{T^{4}/(\mathbb{Z}_2^{0_t} \times \mathbb{Z}_2^{1_p})}(\widetilde{T}'_{+}) =
\begin{pmatrix}
1 & & & \\
 & 1 & & \\
 & & -1 & \\
 & & & -1
\end{pmatrix},
\end{align}
satisfy the following relations,
\begin{align}
\begin{array}{l}
\ \rho_{T^{6}/(\mathbb{Z}_2^{0_t} \times \mathbb{Z}_2^{1_p})}(\widetilde{S})^2 = -iI, \\
\ [ \rho_{T^{6}/(\mathbb{Z}_2^{0_t} \times \mathbb{Z}_2^{1_p})}(\widetilde{S}) \rho_{T^{6}/(\mathbb{Z}_2^{0_t} \times \mathbb{Z}_2^{1_p})}(\widetilde{T}_{I_2}) ]^6 = -iI, \quad 
\ [ \rho_{T^{6}/(\mathbb{Z}_2^{0_t} \times \mathbb{Z}_2^{1_p})}(\widetilde{S}) \rho_{T^{6}/(\mathbb{Z}_2^{0_t} \times \mathbb{Z}_2^{1_p})}(\widetilde{T}_{33}) ]^6 = -I, \\
\ [ \rho_{T^{6}/(\mathbb{Z}_2^{1_t} \times \mathbb{Z}_2^{0_p})}(\widetilde{S}) \rho_{T^{6}/(\mathbb{Z}_2^{0_t} \times \mathbb{Z}_2^{1_p})}(\widetilde{T}_{12}) ]^6 = iI, \quad 
\ [ \rho_{T^{6}/(\mathbb{Z}_2^{1_t} \times \mathbb{Z}_2^{0_p})}(\widetilde{S}) \rho_{T^{6}/(\mathbb{Z}_2^{0_t} \times \mathbb{Z}_2^{1_p})}(\widetilde{T}'_{+}) ]^4 = I, \\
\ \rho_{T^{6}/(\mathbb{Z}_2^{0_t} \times \mathbb{Z}_2^{1_p})}(\widetilde{T}_{I_2}) = \rho_{T^{6}/(\mathbb{Z}_2^{0_t} \times \mathbb{Z}_2^{1_p})}(\widetilde{T}_{12})^{2}, \quad 
\ \rho_{T^{6}/(\mathbb{Z}_2^{0_t} \times \mathbb{Z}_2^{1_p})}(\widetilde{T}'_{+}) = \rho_{T^{4}/(\mathbb{Z}_2^{0_t} \times \mathbb{Z}_2^{1_p})}(\widetilde{T}_{33})^{2}, \\
\ \rho_{T^{6}/(\mathbb{Z}_2^{0_t} \times \mathbb{Z}_2^{1_p})}(\widetilde{T}_{12})^{3} = iI, \\
\ \rho_{T^{6}/(\mathbb{Z}_2^{0_t} \times \mathbb{Z}_2^{1_p})}(\widetilde{T}_{I_2})^3 = \rho_{T^{6}/(\mathbb{Z}_2^{0_t} \times \mathbb{Z}_2^{1_p})}(\widetilde{T}_{12})^{6} = -I, \quad 
\ \rho_{T^{6}/(\mathbb{Z}_2^{0_t} \times \mathbb{Z}_2^{1_p})}(\widetilde{T}'_{+})^2 = \rho_{T^{4}/(\mathbb{Z}_2^{0_t} \times \mathbb{Z}_2^{1_p})}(\widetilde{T}_{33})^{4} = I, \\
\ \rho_{T^{4}/(\mathbb{Z}_2^{0_t} \times \mathbb{Z}_2^{1_p})}(\widetilde{T}_{X}) \rho_{T^{4}/(\mathbb{Z}_2^{0_t} \times \mathbb{Z}_2^{1_p})}(\widetilde{T}_{Y}) =\rho_{T^{4}/(\mathbb{Z}_2^{0_t} \times \mathbb{Z}_2^{1_p})}(\widetilde{T}_{Y}) \rho_{T^{4}/(\mathbb{Z}_2^{0_t} \times \mathbb{Z}_2^{1_p})}(\widetilde{T}_{IX}) \quad (X, Y = I_{2}, 12, 33, +).
\end{array}
\end{align}
Then, they transform non-trivially under $\widetilde{N}_{3,12}^{(3-2-b)}(H)$ modular flavor transformation.
Actually, we can similarly find that the four-dimensional modes transform non-trivially under $(T' \rtimes A_4) \rtimes Z_4$ modular flavor transformation.
If $\tau'_{+}$ is restricted to $\tau'_{+}=0$, they transform non-trivially under $(T' \rtimes Z_2) \times (T' \rtimes Z_4)$ modular flavor transformation, which is nothing but the direct product of $T^4/(\mathbb{Z}_2^{0_t} \times \mathbb{Z}_2^{1_p})$ with the $N$ matrix in Eq.~(\ref{4224}) and $T^2/\mathbb{Z}_2^{0_t}$ with $N=2$.
Therefore, we can obtain the following breaking pattern,
\begin{align}
(T' \rtimes A_4) \rtimes Z_4 \xrightarrow{\tau'_{+}=0} (T' \rtimes Z_2) \times (T' \rtimes Z_4).
\end{align}

Finally, let us see the specific cases of the classes (3-3) and (3-4).
First, let us see the specific case of the class (3-3) with $N_{11}=N_{22}=N_{33}$ and $N_{12}=N_{13}$,~i.e.,
\begin{align}
\begin{array}{l}
N =
\begin{pmatrix}
4 & 2 & 2 \\
2 & 4 & 0 \\
2 & 0 & 4
\end{pmatrix}
= 2
\begin{pmatrix}
2 & 1 & 1 \\
1 & 2 & 0 \\
1 & 0 & 2
\end{pmatrix}
= 2 N', 
\quad
{\rm det}N = 8{\rm det}N' = 32, \quad
\tilde{N} = 4\tilde{N} = 4
\begin{pmatrix}
4 & -2 & -2 \\
-2 & 4 & 1 \\
-2 & 1 & 4
\end{pmatrix}, 
\\
\Omega=
\begin{pmatrix}
\tau+\tau_{N^{-1}} & \tau_{N} & \tau_{N} \\
\tau_{N} & \tau & \tau_{N^{-1}} \\
\tau_{N} & \tau_{N^{-1}} & \tau
\end{pmatrix}
=\tau B_{I_3} + \tau_{N} B_{N_3} + \tau_{N^{-1}} B_{N^{-1}_3}.
\end{array}
\label{422240204}
\end{align}
In this case, the generation number of zero modes on magnetized $T^2/(\mathbb{Z}_2^t \times \mathbb{Z}_2^p)$ with $(m^t,m^p)=(0,1), (1,0), (1,1)$ is six while that of zero modes on magnetized $T^2/(\mathbb{Z}_2^t \times \mathbb{Z}_2^p)$ with $(m^t,m^p)=(0,0)$ is fourteen.
Hence, the modular flavor groups, generated by those dimensional representations of $S$, $T_{I_3}$, $T_{N_3}$, and $T_{N^{-1}_3}$ transformations, become larger and complicated, and then we were not able to specify the modular flavor groups.
Next let us see the specific case of the class (3-4-a) with $N_{11}=N_{22}=N_{33}$ and $N_{12}=N_{23}=N_{31}$,~i.e.,
\begin{align}
\begin{array}{l}
N =
\begin{pmatrix}
4 & 2 & 2 \\
2 & 4 & 2 \\
2 & 2 & 4
\end{pmatrix}
= 2
\begin{pmatrix}
2 & 1 & 1 \\
1 & 2 & 1 \\
1 & 1 & 2
\end{pmatrix}
= 2 N', 
\quad
{\rm det}N = 8{\rm det}N' = 32, \quad
\tilde{N} = 4\tilde{N} = 4
\begin{pmatrix}
3 & -1 & -1 \\
-1 & 3 & -1 \\
-1 & -1 & 3
\end{pmatrix}, 
\\
\Omega=
\begin{pmatrix}
\tau+\tau'_{23} & \tau'_{12} & \tau'_{31} \\
\tau'_{12} & \tau+\tau'_{31} & \tau'_{23} \\
\tau'_{31} & \tau'_{23} & \tau+\tau'_{12}
\end{pmatrix}
=\tau B_{I_3} + \tau'_{12} B'_{12} + \tau'_{23} B'_{23} + \tau'_{31} B'_{31}.
\end{array}
\label{422242224}
\end{align}
In this case, the generation number of $\mathbb{Z}_2^t$ twisted even $(m^t=0)$ modes on magnetized $T^2/\mathbb{Z}_2^t$ is twenty while that of $\mathbb{Z}_2^t$ twisted odd $(m^t=1)$ modes on magnetized $T^2/\mathbb{Z}_2^t$ is twelve.
Hence, the modular flavor groups, generated by those dimensional representations of $S$, $T_{I_3}$, $T'_{12}$, $T'_{23}$, and $T'_{31}$ transformations, become larger and complicated, and then we were not able to specify the modular flavor groups.


\section{Conclusion}
\label{conclusion}

We have studied the modular symmetry in magnetized $T^{2g}$ and orbifold models.
There is $\Gamma_{g}=Sp(2g,\mathbb{Z})$ modular symmetry on $T^{2g}$ and its orbifold by the stabilizer $H$.
When a magnetic flux is introduced on $T^{2g}$ as well as its orbifold, the modular symmetry is reduced from $\Gamma_{g}$ to a certain normalizer $N_{g}(H)$.
We have classified the remaining modular symmetry by magnetic flux matrix types in subsection~\ref{subsec:MSMt2g}.
Furthermore, we have studied modular symmetry for wave functions on the magnetized $T^{2g}$ and certain orbifolds in subsection~\ref{subsec:MSWMT2gOM}.
We have found that wave functions on magnetized $T^{2g}$ as well as its orbifolds behave as the Siegel modular forms of weight $1/2$ and $\widetilde{N}_{g}(H,h)$, which is the metapletic congruence subgroup of the double covering group of $N_{g}(H)$, $\widetilde{N}_{g}(H)$.
Then, they transform non-trivially under the quotient group, $\widetilde{N}_{g,h}=\widetilde{N}_{g}(H)/\widetilde{N}_{g}(H,h)$, where the level $h$ is related to the determinant of the magnetic flux matrix.
Accordingly, the corresponding 4D chiral fields also transform non-trivially under $\widetilde{N}_{g,h}$ modular flavor transformation with modular weight $-1/2$.
We have also studied concrete modular flavor symmetries of wave functions on magnetized $T^{2g}$ orbifold.
The study in this paper is extended from the studies in Refs.~\cite{Kikuchi:2020frp,Kikuchi:2020nxn,Kikuchi:2021ogn} and one specific application of the study in Ref.~\cite{Ding:2020zxw}.

Our results are important to study four-dimensional effective field theory derived by 
torus and orbifold compactifications with magnetic flux background, in particular 
realization of quark and lepton masses.
We would investigate the realistic model building in the magnetized $T^{2g}$ orbifold models elsewhere to understand quark and lepton masses as well as their mixing angles from their modular flavor symmetries.

\vspace{1.5 cm}
\noindent
{\large\bf Acknowledgement}\\

This work was supported by JSPS KAKENHI Grant Numbers JP22KJ0047 (S. K.) and JP23K03375 (T. K.), and JST SPRING Grant Number JPMJSP2119 (K. N. and S. T.)


\appendix


\section{$(ST)^n$ transformation}
\label{ap:STn}

In this Appendix, we discuss the algebraic relations between $S$ transformation and general $T$ transformation generated by $T_{ab}$ transformation.
Generally, $(ST)^n$ can be written as
\begin{align}
(ST)^n =
\begin{pmatrix}
-b_{n-1} & b_{n} \\
-b_{n} & b_{n+1}
\end{pmatrix}, \quad
b_{n+1} = -b_{n}B - b_{n-1}, \label{STnRF}
\end{align}
with
\begin{align}
ST =
\begin{pmatrix}
-b_{0} & b_{1} \\
-b_{1} & b_{2}
\end{pmatrix}
=
\begin{pmatrix}
0 & I_{g} \\
-I_{g} & -B
\end{pmatrix},
\end{align}
where $B$ denotes a general $g \times g$ symmetric matrix.
Here, we assume $\Omega_0$ (${\rm det}\Omega_0 \neq 0$) such that it satisfies
\begin{align}
\Omega_0 + B &= - \Omega_0^{-1}, \label{Omega0} \\
\Leftrightarrow \ 
- (\Omega_0 + B)^{-1} &= \Omega_0, \label{Omega01}
\end{align}
that is, $\Omega_0$ is nothing but the fixed point of $ST$ transformation in moduli space.
Then, the recursion formula in Eq.~(\ref{STnRF}) can be rewritten by
\begin{align}
(b_{n+1} \Omega_0^{n}) - (b_{n} \Omega_0^{n-1}) \Omega_0^2 = I_{g}.
\end{align}
Furthermore, by introducing
\begin{align}
\xi \equiv (I_{g} - \Omega_0^2)^{-1} \quad ({\rm det}(I_{g} \pm \Omega_0) \neq 0 ),
\end{align}
the recursion formula can be solved as
\begin{align}
b_{n} = (I_{g} - \Omega_0^2)^{-1} (I_{g} - \Omega_0^{2n}) \Omega_0^{-(n-1)} = \sum_{k=0}^{n-1} \Omega_0^{2k+1-n},
\label{bn}
\end{align}
where we use the following relation,
\begin{align}
(I_{g} - \Omega_0^2)^{-1} (I_{g} - \Omega_0^{2n}) = \sum_{k=0}^{n-1} \Omega_0^{2k},
\end{align}
for the last equality. 
Now, we can rewrite Eq.~(\ref{STnRF}) as
\begin{align}
(ST)^n =
\begin{pmatrix}
-b_{n}\Omega_0 + \Omega_0^n  & b_{n} \\
-b_{n} & b_{n} \Omega_0^{-1} + \Omega_0^n
\end{pmatrix}, \quad
b_{n} = (I_{g} - \Omega_0^2)^{-1} (I_{g} - \Omega_0^{2n}) \Omega_0^{-(n-1)}.
\end{align}
In particular, when it is satisfied that
$\Omega_0^{2n}=I_{g} \Leftrightarrow b_{n}=0$,
$(ST)^n$ can be written as
\begin{align}
\begin{array}{l}
(ST)^n =
\begin{pmatrix}
\Omega_0^n  & 0 \\
0 & \Omega_0^n
\end{pmatrix} = A_{\Omega_0^n}, \\
(ST)^{2n} = I_{2g}.
\end{array}
\end{align}
Then, in the case that $\Omega_0^{2n+1}=I_{g}$, we can find that $(ST)^{2n+1}=I_{2g}$.

Let us see the meaning of the result in detail.
When we consider $\Omega=\Omega_0$, $(z,\Omega_0)$ transform under $ST$ transformation as
\begin{align}
ST: (z,\Omega_0) \rightarrow (\Omega_0 z, \Omega_0). \label{STtrans}
\end{align}
In addition, if $\Omega_0^n=I_{g} \Leftrightarrow (ST)^n=I_{2g}$, we can find that
\begin{align}
(ST)^n=I_{2g}: (z,\Omega_0) \rightarrow (\Omega_0^n z, \Omega_0)=(z,\Omega_0). \label{STtrans}
\end{align}
This means the $\mathbb{Z}_n$ transformation.

From now, we discuss how to determine the fixed point $\Omega_0$ and the order $n$ such that $(ST)^n=I_{2g}$.
First, the fixed point $\Omega_0$ with $\Omega_0^n=I_{g}$ is written by the diagonalize matrix and a real orthogonal matrix $O$ as
\begin{align}
\begin{array}{c}
\Omega_0 = O^{-1} {\rm diag}(e^{2\pi ik_i/n}) O, \\
{\rm Re}\Omega_0 + i {\rm Im}\Omega_0 = O^{-1} {\rm diag}(\cos (2\pi k_i/n)) O + i O^{-1} {\rm diag}(\sin (2\pi k_i/n)) O.
\end{array}
\end{align}
In addition, from Eq.~(\ref{Omega0}), we can obtain that
\begin{align}
{\rm Re}\Omega_0 = O^{-1} {\rm diag}(\cos (2\pi k_i/n)) O = - \frac{1}{2} B.
\end{align}
Thus, $B$ matrix determines ${\rm Re}\Omega_0$, the eigenvalues, and $O$ matrix.
In addition, from the result, we can also find ${\rm Im}\Omega_0$ and the order $n$.
Here, we note that $B_{ij}=0, \pm 1$ by considering that the fixed point $\Omega_0$ is on the fundamental region, $|(2{\rm Re}\Omega_0)_{ij}| \leq 1$.
In the following, we show the above analysis concretely through examples.

First, let us consider the $g=1$ case. 
\begin{itemize}
\item When $B=1$ ($ST$) case, we obtain ${\rm Re}\Omega_0=-1/2=\cos(2\pi/3)$, and then
\begin{align}
\Omega_0&=e^{2\pi i/3}, \\
\Omega_0^3&=I_{1} \Leftrightarrow (ST)^3=I_{2}. \notag
\end{align}
\item When $B=0$ ($S$) case, we obtain ${\rm Re}\Omega_0=0=\cos(2\pi/4)$, and then
\begin{align}
\Omega_0&=e^{2\pi i/4}, \notag \\
\Omega_0^2&=-I_{1} \Leftrightarrow S^2=-I_{2}, \\
\Omega_0^4&=I_{1} \Leftrightarrow S^4=I_{2}. \notag
\end{align}
\item When $B=-1$ ($ST^{-1}$), we obtain ${\rm Re}\Omega_0=1/2=\cos(2\pi/6)$, and then
\begin{align}
\Omega_0&=e^{2\pi i/6}, \notag \\
\Omega_0^6&=-I_{1} \Leftrightarrow (ST^{-1})^6=-I_{2}, \\
\Omega_0^6&=I_{1} \Leftrightarrow (ST^{-1})^6=I_{2}. \notag
\end{align}
\end{itemize}

Next, let us consider the $g=2$ case.
\begin{itemize}
\item When $B=
\begin{pmatrix}
b_{11} & 0 \\
0 & b_{22}
\end{pmatrix}$ ($ST_{11}^{b_{11}}T_{22}^{b_{22}}$) cases with $b_{ii}=0, \pm 1$, we obtain
\begin{align}
\begin{array}{l}
\Omega_0=
\begin{pmatrix}
e^{2\pi i/n_1} & 0 \\
0 & e^{2\pi i/n_2}
\end{pmatrix}, \quad
n_{i}=\left\{
\begin{array}{ll}
3 & (b_{ii}=1) \\
4 & (b_{ii}=0) \\
6 & (b_{ii}=-1)
\end{array}
\right., \\
\Omega_0^{{\rm lcm}(n_1,n_2)}=I_{2} \Leftrightarrow (ST_{11}^{b_{11}}T_{22}^{b_{22}})^{{\rm lcm}(n_1,n_2)}=I_{4}.
\end{array}
\end{align}
These are nothing but $T^2 \times T^2$ cases. 
\item When $B=
\begin{pmatrix}
1 & \pm1 \\
\pm1 & 1
\end{pmatrix}$ ($ST_{11}T_{22}T_{12}^{\pm1}$) and $B=
\begin{pmatrix}
-1 & \pm1 \\
\pm1 & -1
\end{pmatrix}$ ($ST_{11}^{-1}T_{22}^{-1}T_{12}^{\pm1}$) cases, we obtain
\begin{align}
\Omega_0=O^{-1}
\begin{pmatrix}
e^{2\pi i/4} & 0 \\
0 & \pm1
\end{pmatrix}O.
\end{align}
From it, however, there is no solution since ${\rm det}(I-\Omega_0^2)=0$.
\item When $B=
\begin{pmatrix}
0 & \pm1 \\
\pm1 & 0
\end{pmatrix}$ ($ST_{12}^{\pm1}$) case, we obtain
\begin{align}
\Omega_0
&=O^{-1}
\begin{pmatrix}
e^{2\pi i/3} & 0 \\
0 & e^{2\pi i/6}
\end{pmatrix}O\notag \\
&=
\begin{pmatrix}
\frac{\sqrt{3}}{2} i & \mp \frac{1}{2} \\
\mp \frac{1}{2} & \frac{\sqrt{3}}{2} i
\end{pmatrix}, \\
\Omega_0^3&=\pm B_{12} \Leftrightarrow (ST_{12}^{\pm1})^3=\pm A_{B_{12}}, \notag \\
\Omega_0^6&=I_{2} \Leftrightarrow (ST_{12}^{\pm1})^6=I_{4}. \notag
\end{align}
\item When $B=
\begin{pmatrix}
1 & \pm1 \\
\pm1 & -1
\end{pmatrix}$ ($ST_{11}T_{22}^{-1}T_{12}^{\pm1}$) case, we obtain 
\begin{align}
\Omega_0&=O^{-1}
\begin{pmatrix}
e^{2\pi i/8} & 0 \\
0 & e^{6\pi i/8}
\end{pmatrix}O \notag \\
&=
\begin{pmatrix}
-\frac{1}{2}+\frac{i}{\sqrt{2}} & \mp \frac{1}{2} \\
\mp \frac{1}{2} & \frac{1}{2}+\frac{i}{\sqrt{2}}
\end{pmatrix}, \\
\Omega_0^4 &=-I_{2} \Leftrightarrow (ST_{11}T_{22}^{-1}T_{12}^{\pm1})^4=-I_{4}, \notag \\
\Omega_0^8 &=I_{2} \Leftrightarrow (ST_{11}T_{22}^{-1}T_{12}^{\pm1})^8=I_{4}. \notag 
\end{align}
Similarly, we ca find that $(ST_{11}^{-1}T_{22}T_{12}^{\pm1})^5=I_{4}$.
\item When $B=
\begin{pmatrix}
1 & \pm1 \\
\pm1 & 0
\end{pmatrix}$ ($ST_{11}T_{12}^{\pm1}$) case, we obtain 
\begin{align}
\Omega_0&=O^{-1}
\begin{pmatrix}
e^{2\pi i/5} & 0 \\
0 & e^{4\pi i/5}
\end{pmatrix}O \notag \\
&=
\begin{pmatrix}
-\frac{1}{2}+\frac{i}{2}\left( \sqrt{\frac{5+\sqrt{5}}{10}} + \sqrt{\frac{5-\sqrt{5}}{10}} \right) & \mp \left( \frac{1}{2}+\frac{i}{2}\left( \sqrt{\frac{5+\sqrt{5}}{10}} -  \sqrt{\frac{5-\sqrt{5}}{10}} \right) \right) \\
\mp \left( \frac{1}{2}+\frac{i}{2}\left( \sqrt{\frac{5+\sqrt{5}}{10}} -  \sqrt{\frac{5-\sqrt{5}}{10}} \right) \right) & \frac{i}{2}\left( \sqrt{\frac{5+2\sqrt{5}}{5}} + \sqrt{\frac{5-2\sqrt{5}}{5}} \right)
\end{pmatrix}, \\
\Omega_0^5 &=I_{2} \Leftrightarrow (ST_{11}T_{12}^{\pm1})^5=I_{4}. \notag
\end{align}
Similarly, we can find that $(ST_{22}T_{12}^{\pm1})^5=I_{4}$.
\item When $B=
\begin{pmatrix}
-1 & \pm1 \\
\pm1 & 0
\end{pmatrix}$ ($ST_{11}^{-1}T_{12}^{\pm1}$) case, we obtain 
\begin{align}
\Omega_0&=O^{-1}
\begin{pmatrix}
e^{2\pi i/10} & 0 \\
0 & e^{6\pi i/10}
\end{pmatrix}O \notag \\
&=
\begin{pmatrix}
\frac{1}{2}+\frac{i}{2}\left( \sqrt{\frac{5+\sqrt{5}}{10}} + \sqrt{\frac{5-\sqrt{5}}{10}} \right) & \mp \left( \frac{1}{2}-\frac{i}{2}\left( \sqrt{\frac{5+\sqrt{5}}{10}} -  \sqrt{\frac{5-\sqrt{5}}{10}} \right) \right) \\
\mp \left( \frac{1}{2}-\frac{i}{2}\left( \sqrt{\frac{5+\sqrt{5}}{10}} -  \sqrt{\frac{5-\sqrt{5}}{10}} \right) \right) & \frac{i}{2}\left( \sqrt{\frac{5+2\sqrt{5}}{5}} + \sqrt{\frac{5-2\sqrt{5}}{5}} \right)
\end{pmatrix}, \\
\Omega_0^{5} &=-I_{2} \Leftrightarrow (ST_{11}^{-1}T_{12}^{\pm1})^{5}=-I_{4}, \notag \\
\Omega_0^{10} &=I_{2} \Leftrightarrow (ST_{11}^{-1}T_{12}^{\pm1})^{10}=I_{4}. \notag
\end{align}
Similarly, we can find that $(ST_{22}^{-1}T_{12}^{\pm1})^{10}=I_{4}$.
\end{itemize}

Finally, let us consider the $g=3$ case.
\begin{itemize}
\item When $B=
\begin{pmatrix}
b_{11} & b_{12} & 0 \\
b_{12} & b_{22} & 0 \\
0 & 0 & b_{33}
\end{pmatrix}$ ($ST_{11}^{b_{11}}T_{22}^{b_{22}}T_{33}^{b_{33}}T_{12}^{b_{12}}$),
$B=
\begin{pmatrix}
b_{11} & 0 & b_{13} \\
0 & b_{22} & 0 \\
b_{13} & 0 & b_{33}
\end{pmatrix}$ ($ST_{11}^{b_{11}}T_{22}^{b_{22}}T_{33}^{b_{33}}T_{13}^{b_{13}}$), and
$B=
\begin{pmatrix}
b_{11} & 0 & 0 \\
0 & b_{22} & b_{23} \\
0 & b_{23} & b_{33}
\end{pmatrix}$ ($ST_{11}^{b_{11}}T_{22}^{b_{22}}T_{33}^{b_{33}}T_{23}^{b_{23}}$) cases with $b_{ij}=0, \pm 1$, they are nothing but $T^4 \times T^2$ cases.
\item When $B=
\begin{pmatrix}
0 & 0 & \pm_1 1 \\
0 & 0 & \pm_2 1 \\
\pm_1 1 & \pm_2 1 & 0
\end{pmatrix}$ ($ST_{13}^{\pm_1 1}T_{23}^{\pm_2 1}$) case, we obtain
\begin{align}
\Omega_0&=O^{-1}
\begin{pmatrix}
e^{6\pi i/8} & 0 & 0 \\
0 & e^{2\pi i/8} & 0 \\
0 & 0 & e^{4\pi i/8}
\end{pmatrix}O\notag \\
&=
\begin{pmatrix}
\left( \frac{1}{2} + \frac{1}{2\sqrt{2}} \right) i & (\pm_1 1)(\pm_2 1) \left( -\frac{1}{2} + \frac{1}{2\sqrt{2}} \right) i & \mp_1 \frac{1}{2} \\
 (\pm_1 1)(\pm_2 1) \left( -\frac{1}{2} + \frac{1}{2\sqrt{2}} \right) i & \left( \frac{1}{2} + \frac{1}{2\sqrt{2}} \right) i & \mp_2 \frac{1}{2} \\
\mp_1 \frac{1}{2} & \mp_2 \frac{1}{2} & \frac{i}{\sqrt{2}}
\end{pmatrix}, \\
\Omega_0^{4} &=-((\pm_1 1)(\pm_2 1)B_{12}+B_{33}) \Leftrightarrow (ST_{13}^{\pm_1 1}T_{23}^{\pm_2 1})^{4}=-A_{(\pm_1 1)(\pm_2 1)B_{12}+B_{33}}, \notag \\
\Omega_0^{8} &=I_{3} \Leftrightarrow (ST_{13}^{\pm_1 1}T_{23}^{\pm_2 1})^{8}=I_{6}. \notag
\end{align}
Similarly, we can find that $(ST_{12}^{\pm_1 1}T_{23}^{\pm_2 1})^{8}=(ST_{12}^{\pm_1 1}T_{13}^{\pm_2 1})^{8} =I_{6}$.
\end{itemize}


\section{Landsberg-Schaar relation}
\label{ap:Landsberg-Schaar}

We derive the $g$-dimensional Landsberg-Schaar relation with $N$ and $B$ matrices, satisfying
\begin{align}
\begin{array}{l}
N_{ij}=N_{ji}, \\
B_{ij}=B_{ji}, \\
(NB)_{ij}=(NB)_{ji}, \\
(NB)_{ii} \in 2\mathbb{Z},
\end{array}
\end{align}
where $i, j \in \{ 1,2, \cdots,g \}$.
In the process of derivation of the Landsberg-Schaar relation, we use the Poisson resummation formula, and to make it converge, we introduce an infinitesimal positive number $\epsilon$ ($0<\epsilon << 1$). 

First, we introduce the following function $f(A)$ to describe the Poisson resummation formula with $g\times g$ symmetric matrix $A$,
\begin{align}
f(A)=\sum_{K\in \mathbb{Z}^{g}} e^{-\pi {^t}K A K}.
\end{align}
We find the Poisson resummation formula,
\begin{align}
f(A)=\frac{1}{\sqrt{\det A}} f(A^{-1}). 
\end{align}

Next, we define $A^{-1}$ as
\begin{align}
A^{-1}= i B^{-1} N + \epsilon {I}_{g}.
\end{align}
We will take the limit $\epsilon \rightarrow +0$ later to obtain the Landsberg-Schaar relation.
Hereafter, we ignore higher-order terms of $\epsilon$ because the final result we obtain is unaffected.
Then, $A$ can be written as
\begin{align}
A &= -iBN^{-1} + \epsilon B^2 N^{-2}.
\end{align}

Now, $f(A)$ can be written as
\begin{align}
f(A)
&=\sum_{K\in \mathbb{Z}^{g}} e^{-\pi {^t}K (-i B N^{-1} + \epsilon B^2 N^{-2}) K} \notag\\
&=\sum_{K\in \mathbb{Z}^{g}}e^{- \pi \epsilon {^t}(B N^{-1}K) ( B N^{-1}K)} e^{\pi i {^t}K B N^{-1}K} \notag\\
&= \sum_{L \in \Lambda_N} \sum_{J \in \mathbb{Z}^{g}}  e^{-\pi \epsilon ({^t}J B + {^t}L B N^{-1} ) (B J+ B N^{-1} L)} e^{\pi i ({^t}J N + {^t}L ) B N^{-1} (N J + L)} \notag\\
&= \sum_{L \in \Lambda_N} e^{\pi i {^t}L B N^{-1} L} \sum_{J \in \mathbb{Z}^{g}}  e^{-\pi \epsilon {^t}(J + N^{-1} L) B^2 (J + N^{-1} L)},
\end{align}
where we write $K=NJ + L$ ($J \in \mathbb{Z}^{g} ,L \in \Lambda_{N}$) in the third equality, and we use $(NB)_{ii} \in 2\mathbb{Z}$ in the fourth equality.
In the limit $\epsilon \rightarrow +0$, we can obtain
\begin{align}
\lim_{\epsilon \rightarrow +0} f(A)
= \lim_{\epsilon \rightarrow +0}
\frac{1}{|\det{B}|} \frac{1}{\epsilon ^{g/2}} \sum_{L\in \Lambda_N} e^{\pi i {^t}L B N^{-1}L},
\end{align}
where we use the formula of Gaussian integral with multiple-variables.
On the other hand, $f(A^{-1})/\sqrt{\det{A}}$ can be written as
\begin{align}
&\frac{1}{\sqrt{\det A}}f(A^{-1}) \notag \\
&=\frac{1}{\sqrt{\det A}} \sum_{K \in \mathbb{Z}^{g}} e^{-\pi {^t}K (i B^{-1}N + \epsilon {I}_{g} ) K } \notag\\
&=\frac{1}{\sqrt{\det A}} \sum_{K \in \mathbb{Z}^{g}} e^{-\pi \epsilon {^t}K K} e^{-\pi i {^t}K B^{-1} N K} \notag\\
&=\frac{1}{\sqrt{\det A}} \sum_{L \in \Lambda_B} \sum_{J \in \mathbb{Z}^{g}} 
e^{-\pi \epsilon {^t}(B J+ L) (BJ+ L)}e^{-\pi i {^t}(B J + L) B^{-1} N (B J + L) } \notag\\
&=\frac{1}{\sqrt{\det A}} \sum_{L\in \Lambda_B} e^{-\pi i {^t}L B^{-1} N L } \sum_{J \in \mathbb{Z}^{g}} 
e^{- \pi \epsilon {^t}(J+ B^{-1}L) B^{2} (J + B^{-1}L)},
\end{align}
where we write $K= B J+ L$ ($J \in \mathbb{Z}^{g}$, $L \in \Lambda_{B}$) in the third equality, and we use $(NB)_{ii} \in 2\mathbb{Z}$ in the fourth equality.
In the limit $\epsilon \rightarrow +0$, we can obtain
\begin{align}
\lim_{\epsilon \rightarrow +0} \frac{1}{\sqrt{\det A}}f(A^{-1})
= \lim_{\epsilon \rightarrow +0}
\sqrt{|\det N|} \frac{e^{\pi i(g+2(n^N_{-}-n^B_{-}))/4}}{\sqrt{|\det{B}}|}
\frac{1}{|\det{B}|} \frac{1}{\epsilon^{g/2}}\sum_{L\in \Lambda_B} e^{-\pi i {^t}L B^{-1} N L},
\end{align}
where we use the formula of Gaussian integral with multiple-variables and we take $(\pm i)^{x/2}=e^{\pm \pi i x/4}$ with the number of negative eigenvalues of $B$ ($N$) matrix, $n^B_{-}$ ($n^N_{-}$).

Thus, we can obtain the $g$-dimensional Landsberg-Schaar relation:
\begin{align}
\frac{e^{-\pi i n^N_{-}/2}}{\sqrt{|{\rm det}N}|} \sum_{K \in \Lambda_N} e^{\pi i {^t}K N^{-1}B K} = \frac{e^{\pi ig/4}e^{-\pi i n^{B}_{-}/2}}{\sqrt{|{\rm det}B}|} \sum_{K \in \Lambda_B} e^{-\pi i {^t}K NB^{-1} K}. \label{LSrelorg}
\end{align}
In Eq.~(\ref{LSrel}), we assume that all eigenvalues of $N$ matrix are positive,~i.e. $n^N_{-}=0$.


\section{The generators of $\Delta(96) \times Z_4$}
\label{ap:proof}

In this Appendix, we prove the generators in Eq.~(\ref{Delta96Z4gen}) satisfy the algebraic relations of $\Delta(96) \times Z_4$ in Eq.~(\ref{Delta96Z4alg}), where $S$, $T_{I_2}$, and $T_{12}$ satisfy the following relations,
\begin{align}
\begin{array}{l}
T_{12}^2=1, \\
(ST_{12})^3=-i1, \\
(ST_{12})^6=S^2=-1, \\
(ST_{12})^{12}=S^4=(ST_{I_2})^3=1, \\
T_{I_2}^4=1.
\end{array}
\end{align}
In addition, from Eq.~(70) in Ref.~\cite{Kikuchi:2021ogn}, the above Eq.~(\ref{2002rho3}) also satisfies
\begin{align}
1
&= (S^{-1}T_{I_2}^{-1}ST_{I_2})^3 \notag \\
&= (T_{I_2}ST_{I_2}SST_{I_2})^3 \notag \\
&= (T_{I_2}S^3T_{I_2}^2)^3 \notag \\
&= (T_{I_2}^2S^3T_{I_2})^3 \notag \\
&=(S^3T_{I_2}^3).
\end{align}
We note that, in Ref.~\cite{Kikuchi:2021ogn}, we have already proved that the generators in Eq.~(\ref{S'4gen}) satisfy the algebraic relation of $S'_4 \simeq \Delta'(24)$ in Eq.~(\ref{S'4alg}).
First, we can easily check that $d=(ST_{12})^3$ satisfies
\begin{align}
&d^4=1, \\
&dx = xd\ (x=\tilde{a}, \tilde{a}', b, c').
\end{align}
Next, $\tilde{a}'$ can be rewritten as
\begin{align}
\tilde{a}'
&= ST_{I_2}T_{12}S^{-1}T_{12}^{-1}T_{I_2}^{-1} \\
&= ST_{I_2} ST_{12}S^{-1}T_{I_2}^{-1}(ST_{12})^3 \notag \\
&= T_{I_2}^{-1}S^{-1}T_{I_2}^{-1}T_{12}T_{I_2}ST_{I_2}S (ST_{12})^3 \notag \\
&= T_{I_2}^{-1}ST_{12}ST_{I_2}S^{-1} (ST_{12})^3 \notag \\
&= T_{I_2}^{-1} T_{12}^{-1} ST_{12}T_{I_2}S^{-1} \notag \\
&= T_{12}^{-1}T_{I_2}^{-1} ST_{I_2}T_{12}S^{-1},
\end{align}
and then we can prove that
\begin{align}
\tilde{a}'^2
&= ST_{I_2}T_{12}S^{-1}T_{12}^{-1}T_{I_2}^{-1} ST_{I_2}T_{12}S^{-1}T_{12}^{-1}T_{I_2}^{-1} \notag \\
&= ST_{I_2}T_{12}S^{-1} ST_{I_2}T_{12}S^{-1} T_{12}^{-2} T_{I_2}^{-2} \notag \\
&= ST_{I_2}^2S^{-1} T_{I_2}^{-2}. \notag \\
&= a'.
\end{align}
Similarly, $\tilde{a}$ can be rewritten as
\begin{align}
\tilde{a}
&= T_{I_2}T_{12} ST_{12}T_{I_2}S^{-1} T_{I_2}T_{12} (ST_{12})^3 \notag \\
&= T_{I_2}^2 ST_{12}T_{I_2}S^{-1} (ST_{12})^3 \notag \\
&= ST_{12}T_{I_2}S^{-1} T_{I_2}^2 (ST_{12})^3 ,
\end{align}
and then we can prove that
\begin{align}
\tilde{a}^2
&= T_{I_2}^2 ST_{12}T_{I_2}S^{-1} (ST_{12})^3 ST_{12}T_{I_2}S^{-1} T_{I_2}^2 (ST_{12})^3 \notag \\
&= T_{I_2}^2 ST_{I_2}^2S T_{I_2}^2 \notag \\
&= a.
\end{align}
They are the proof of the Eq.~(\ref{atildea}).
Hence, by considering Eq.~(\ref{S'4alg}), we can obtain that
\begin{align}
&\tilde{a}^4=1, \\
&\tilde{a}'^4=1.
\end{align}
We can also check that
\begin{align}
\tilde{a}\tilde{a'} = \tilde{a}'\tilde{a} = ST_{I_2}^2S^{-1} T_{I_2}T_{12} (ST_{12})^3.
\end{align}
Similarly, since $b$ in Eq.~(\ref{Delta96Z4gen}) and $b$ in Eq.~(\ref{S'4gen}) are same, we can obtain that
\begin{align}
b^3=1.
\end{align}
On the other hand, by considering that $c$ in Eq.~(\ref{S'4gen}) satisfies that
\begin{align}
c^2 = S^2,
\end{align}
we can obtain that
\begin{align}
c'^2=1.
\end{align}
We also obtain
\begin{align}
c'bc'^{-1} = cbc^{-1} = b^{-1}.
\end{align}
The other relations can be checked that
\begin{align}
b\tilde{a}b^{-1}
&= T_{I_2}S^3T_{I_2}^2 ST_{12}T_{I_2}S^{-1} T_{I_2}^2 T_{I_2}^{-2}S^{-3}T_{I_2}^{-1} (ST_{12})^3 \notag \\
&= S^{-1}T_{I_2}^{-1}S T_{I_2} ST_{12} (ST_{12})^3 \notag \\
&= S^{-1} T_{I_2}^{-2} S^{-1} T_{I_2}^{-1}T_{12}^{-1} (ST_{12})^3 \notag \\
&= (ST_{12})^{-3} T_{12}^{-1} T_{I_2}^{-1} ST_{I_2}^{-2}S^{-1} \notag \\
&= \tilde{a}^{-1} \tilde{a}'^{-1}, \\
b\tilde{a}'b^{-1}
&= T_{I_2}S^3T_{I_2}^2 T_{I_2}^{-1}T_{12}^{-1} ST_{12}T_{I_2}S^{-1} T_{I_2}^{-2}S^{-3}T_{I_2}^{-1} \notag \\
&= T_{I_2}ST_{I_2} T_{12}ST_{12} T_{I_2}^2 T_{I_2}^{-1}S^{-1}T_{I_2}^{-1} T_{I_2}^{-1}S^{-1}T_{I_2}^{-1} \notag \\
&= S^{-1}T_{I_2}^{-1}S^{-1} S^{-1}T_{12}^{-1}S^{-1} T_{I_2}^2 ST_{I_2}S ST_{I_2}S S^{-2} (ST_{12})^{-3} \notag \\
&= S^{-1}T_{I_2}^{-1}T_{12}^{-1}S^{-1} T_{I_2}^2 ST_{I_2}^2S (ST_{12})^3 \notag \\
&= S^{-1} T_{12} T_{I_2}^{-1} S^{-1} ST_{I_2}^2S T_{I_2}^2 (ST_{12})^3 \notag \\
&= ST_{12}T_{I_2}S^{-1} T_{I_2}^2 (ST_{12})^3 \notag \\
&= \tilde{a}, \\
c' \tilde{a} c'^{-1}
&= ST_{I_2}^2ST_{I_2}^3 ST_{I_2}T_{12}S^{-1}T_{I_2}^2 T_{I_2}^{-3}S^{-1}T_{I_2}^{-2}S^{-1} (ST_{12})^3 \notag \\
&= S T_{I_2}^2S^3T_{I_2} T_{I_2}^2S^3T_{I_2} T_{12} S^{-1}T_{I_2}^{-1}S^{-1}T_{I_2}^{-1} T_{I_2}^{-1}S^{-1} (ST_{12})^3 \notag \\
&= S T_{I_2}^{-1}S^{-3}T_{I_2}^{-2} T_{12} T_{I_2}S ST_{I_2}ST_{I_2} (ST_{12})^3 \notag \\
&= S^{-1}T_{I_2}^{-1}ST_{12}ST_{I_2} (ST_{12})^3 \notag \\
&= S^{-1}T_{I_2}^{-1} T_{12}^{-1}ST_{12}T_{I_2} \notag \\
&= T_{I_2}T_{12}ST_{12}^{-1}T_{I_2}^{-1}S^{-1} \notag \\
&= \tilde{a}'^{-1}, \\
c' \tilde{a}' c'^{-1}
&= ST_{I_2}^2ST_{I_2}^3 T_{I_2}^{-1}T_{12}^{-1}ST_{I_2}T_{12}S^{-1} T_{I_2}^{-3}S^{-1}T_{I_2}^{-2}S^{-1} \notag \\
&= T_{I_2}^{-1}T_{12}^{-1} S T_{I_2}^2S^3T_{I_2} T_{I_2}^2S^3T_{I_2} T_{12}T_{I_2} T_{I_2}^{-1}S^{-3}T_{I_2}^{-2} T_{I_2}^{-1}S^{-3}T_{I_2}^{-2} S^{-1} \notag \\
&= T_{I_2}^{-1}T_{12}^{-1} S T_{I_2}^{-1}S^{-3}T_{I_2}^{-2} T_{12}T_{I_2} T_{I_2}^2S^3T_{I_2} S^{-1} \notag \\
&= T_{12}^{-1} T_{I_2}^{-1}S^{-1}T_{I_2}^{-1}S^{-1}T_{I_2}^{-1} T_{12} T_{I_2}^2S^3T_{I_2} S^{-1} \notag \\
&= T_{12}ST_{12} T_{I_2}^2S^3T_{I_2} S^{-1} \notag \\
&= ST_{12}S T_{I_2}^2ST_{I_2}S^{-1} (ST_{12})^{-3} \notag \\
&= ST_{12}^{-1}T_{I_2}^{-1} T_{I_2}ST_{I_2} T_{I_2}ST_{I_2} S^{-1} (ST_{12})^{-3} \notag \\
&= ST_{12}^{-1}T_{I_2}^{-1} S^{-1}T_{I_2}^{-2} \notag \\
&= \tilde{a}^{-1},
\end{align}
where we also used the following relation proved in Ref.~\cite{Kikuchi:2021ogn},
\begin{align}
ST_{I_2}^{2p}S^{-1}T_{I_2}^{2q} =(ST_{I_2}^2S^{-1})^{p}T_{I_2}^{2q} = T_{I_2}^{2q}ST_{I_2}^{2p}S^{-1}, \quad p,q \in \mathbb{Z}. \label{T2p2q}
\end{align}
Therefore, the generators in Eq.~(\ref{Delta96Z4gen}) satisfy the algebraic relations of $\Delta(96) \times Z_4$ in Eq.~(\ref{Delta96Z4alg}).


\section{Modular flavor symmetry with $\Omega=\tau I_{g}$ case}
\label{ap:OmegatauI}

In this Appendix, let us see the modular flavor symmetry of wave functions on magnetized $T^{2g}$ and its orbifolds in particular for $\Omega=\tau I_{g}$ case.
First, in $\Omega=\tau I_{g}$ case, we can consider $\Gamma_{1}=SL(2,\mathbb{Z})$ modular transformation for any $N$ matrices.
In addition, in order for the modular transformation to be consistent with the boundary conditions of wave functions on magnetized $T^{2g}$ as well as its orbifolds, it is required that the diagonal elements of $N$ matrix must be even ($N_{ii} \in 2\mathbb{Z}$) in the case of the vanishing SS phases.
In this case, the wave functions on magnetized $T^{2g}$ transform under the modular transformation as
\begin{align}
&\widetilde{\Gamma}_{1} \ni \widetilde{\gamma} : \psi^{J,N}_{T^{2g}}(z,\Omega) \rightarrow \psi^{J,N}_{T^{2g}}(\widetilde{\gamma}(z,\Omega)) = \widetilde{J}_{g/2}(\widetilde{\gamma},\Omega) \rho_{T^{2g}}(\widetilde{\gamma})_{JK} \psi^{K,N}_{T^{2g}}(z,\Omega), \label{modularwavetauI} \\
&\widetilde{S} = [S,(-1)^g] : \quad \widetilde{J}_{g/2}(\widetilde{S},\Omega) = (-1)^g (-\tau)^{g/2}, \quad \rho_{T^{2g}}(\widetilde{S})_{JK} = \frac{(-e^{\pi i/4})^g}{\sqrt{{\rm det}N}} e^{2\pi i{^t}JN^{-1}K}, \label{StildetauI} \\
&\widetilde{T}_{I_{g}} = [T_{I_g},1] :  \quad \widetilde{J}_{g/2}(\widetilde{T}_{I_g},\Omega) = 1, \quad \rho_{T^{2g}}(\widetilde{T}_{I_g})_{JK} = e^{\pi i{^t}JN^{-1}J} \delta_{J,K}, \label{TtildetauI}
\end{align}
where $\rho_{T^{2g}}$ satisfy the algebraic relations in Eq.~(\ref{Stildewaverel}), the top of Eq.~(\ref{ST3tildewaverel}) with $B=I_{g}$ and $n^B_{-}=0$, and Eq.~(\ref{Ttildewaverel}).
Thus, the wave functions behave as the modular forms of weight $g/2$ and $\widetilde{\Gamma}_{1}(h)$, and then they transform under $\widetilde{\Gamma}_{1,h}=\widetilde{\Gamma}_{1}/\widetilde{\Gamma}_{1}(h)$ modular flavor transformation non-trivially.
Note that, for $g=2$, it corresponds to $\Gamma'_{1,h}=\Gamma_{1}/\Gamma_{1}(h)$ modular flavor transformation.
From now, we show concrete modular flavor symmetries of three-dimensional wave functions.
$g=1$ cases have been studied in Refs.~\cite{Kikuchi:2020frp,Kikuchi:2021ogn}.
Once we study $g=2$ cases, we can similarly study $g=3$ case and the result is similar to $g=1$ cases by replacing the modular weight from $1/2$ to $3/2$.
Hence, we show examples of $g=2$ cases, in particular.
Moreover, the cases of the class (2-1) have been studied in Refs.~\cite{Kikuchi:2020nxn,Kikuchi:2021ogn}. 
Then, let us see the case of the class (2-2).

First, let us see the case of the class (2-2-b).
Ref.~\cite{Kikuchi:2022lfv} shows that only $\mathbb{Z}_2^t$ twisted odd ($m^t=1$) modes with ${\rm det}N=7$ are three-dimensional modes on magnetized $T^{4}/\mathbb{Z}_2^t$ with vanishing SS phases and $N_{ii} \in 2\mathbb{Z}$.
In this case, we can find that $s=1$ and $\tilde{N}'_{ii}=\tilde{N}_{ii} \in 2\mathbb{Z}$.
Hence, the order $h$ is determined as $h={\rm det}N=7$.
In addition, by considering that the wave functions are $\mathbb{Z}_2^t$ twisted odd modes, we obtain the following algebraic relations,
\begin{align}
\begin{array}{l}
\ \rho_{T^{4}/\mathbb{Z}_2^{1}}(\widetilde{S})^2 = I, \\
\ [\rho_{T^{4}/\mathbb{Z}_2^{1}}(\widetilde{S}) \rho_{T^{4}/\mathbb{Z}_2^{1}}(\widetilde{T}_{I_2})]^3 = I, \\
\ \rho_{T^{4}/\mathbb{Z}_2^{1}}(\widetilde{T}_{I_2})^7 = I.
\end{array}
\label{rho3relGamma17}
\end{align}
Thus, the three-dimensional modes transform non-trivially under $\Gamma_{1,7}=PSL(2,\mathbb{Z}_7)$ modular flavor transformation.
Indeed, let us see the following example,
\begin{align}
\begin{array}{l}
N =
\begin{pmatrix}
2 & 1 \\
1 & 4
\end{pmatrix}
=N', \quad
{\rm det}N^{(\prime)} = 7, \quad
\tilde{N}^{(\prime)} = 
\begin{pmatrix}
4 & -1 \\
-1 & 2
\end{pmatrix}. \\
\end{array}
\label{2114}
\end{align}
The three-dimensional $\mathbb{Z}_2^t$ twisted odd ($m^t=1$) modes,
\begin{align}
\ket{J_1,J_2} =
\begin{pmatrix}
\frac{1}{\sqrt{2}} (\ket{1,1}-\ket{2,4}) \\  
\frac{1}{\sqrt{2}} (\ket{1,2}-\ket{2,3}) \\ 
\frac{1}{\sqrt{2}} (\ket{1,3}-\ket{2,2})
\end{pmatrix}
\label{2114gen}
\end{align}
transform under $S$ and $T_{I_2}$ transformations as
\begin{align}
\rho_{T^{4}/\mathbb{Z}_2^1}(\widetilde{S}) &= \frac{2}{\sqrt{7}}
\begin{pmatrix}
\sin(\frac{6\pi}{7}) &  \sin(\frac{4\pi}{7}) &  \sin(\frac{2\pi}{7}) \\
\sin(\frac{4\pi}{7}) & - \sin(\frac{2\pi}{7}) &  \sin(\frac{6\pi}{7})\\
\sin(\frac{2\pi}{7}) &  \sin(\frac{6\pi}{7}) & - \sin(\frac{4\pi}{7})
\end{pmatrix}, \quad
\rho_{T^{4}/\mathbb{Z}_2^1}(\widetilde{T}_{I_2}) &=
\begin{pmatrix}
e^{4\pi i/7} & & \\
 & e^{8\pi i/7} & \\
 & & e^{2\pi i/7}
\end{pmatrix},
\label{rho3Gamma17}
\end{align}
which satisfy the algebraic relations in Eq.~(\ref{rho3relGamma17}) and also
\begin{align}
\left[ \rho_{T^{4}/\mathbb{Z}_2^1}(\widetilde{S})^{-1}\rho_{T^{4}/\mathbb{Z}_2^1}(\widetilde{T}_{I_2})^{-1}\rho_{T^{4}/\mathbb{Z}_2^1}(\widetilde{S})\rho_{T^{4}/\mathbb{Z}_2^1}(\widetilde{T}_{I_2}) \right]^4 = I.
\label{S-1T-1ST4}
\end{align}
Thus, they transform non-trivially under $\Gamma_{1,7}=PSL(2,\mathbb{Z}_7)$ modular flavor transformation.

Next, let us see the case of the class (2-2-a).
There are two types of examples, besides one in Eqs.~(\ref{4224})-(\ref{Delta96Z4Z3S3Z3}), such that there appear three-dimensional modes on magnetized $T^4/(\mathbb{Z}_2^t \times \mathbb{Z}_2^p)$ with vanishing SS phases and $N_{11}=N_{22}=n \in 2\mathbb{Z}$.
The first one is that the $N$ matrix is given by
\begin{align}
\begin{array}{l}
N =
\begin{pmatrix}
4 & 3 \\
3 & 4
\end{pmatrix}
=N', \quad
{\rm det}N^{(\prime)} = 7, \quad
\tilde{N}^{(\prime)} = 
\begin{pmatrix}
4 & -3 \\
-3 & 4
\end{pmatrix}. \\
\end{array}
\label{4334}
\end{align}
The three-dimensional $\mathbb{Z}_2^t$ twisted odd ($m^t=1$) and $\mathbb{Z}_2^p$ permutation even ($m^p=0$) modes,\footnote{There are no $\mathbb{Z}_2^t$ twisted odd and $\mathbb{Z}_2^p$ permutation odd modes. In other words, all $\mathbb{Z}_2^t$ twisted odd modes are $\mathbb{Z}_2^p$ permutation even.}
\begin{align}
\ket{J_1,J_2} =
\begin{pmatrix}
\frac{1}{\sqrt{2}} (\ket{3,3}-\ket{4,4}) \\
\frac{1}{\sqrt{2}} (\ket{|2,2}-\ket{5,5}) \\ 
\frac{1}{\sqrt{2}} (\ket{1,1}-\ket{6,6})
\end{pmatrix}
\label{4334gen}
\end{align}
also transform under $S$ and $T_{I_2}$ transformations as Eqs.~(\ref{rho3Gamma17}) and (\ref{S-1T-1ST4}), and then  they also transform non-trivially under $\Gamma_{1,7}=PSL(2,\mathbb{Z}_7)$ modular flavor transformation.
On the other hand, the second one is that the $N$ matrix is given by
\begin{align}
\begin{array}{l}
N =
\begin{pmatrix}
4 & 1 \\
1 & 4
\end{pmatrix}
=N', \quad
{\rm det}N^{(\prime)} = 15, \quad
\tilde{N}^{(\prime)} = 
\begin{pmatrix}
4 & -1 \\
-1 & 4
\end{pmatrix}. \\
\end{array}
\label{4114}
\end{align}
The three-dimensional $\mathbb{Z}_2^t$ twisted odd ($m^t=1$) and $\mathbb{Z}_2^p$ permutation odd ($m^p=1$) modes, 
\begin{align}
\ket{J_1,J_2} =
\begin{pmatrix}
\frac{1}{2} (\ket{2,1}-\ket{1,2}-\ket{3,4}+\ket{4,3}) \\
\frac{1}{2} (\ket{3,1}-\ket{1,3}-\ket{2,4}+\ket{4,2}) \\
\frac{1}{\sqrt{2}} (\ket{3,2}-\ket{2,3})
\end{pmatrix},
\label{4114gen}
\end{align}
transform under $S$ and $T_{I_2}$ transformations as
\begin{align}
\rho_{T^{4}/(\mathbb{Z}_2^{1_t} \times \mathbb{Z}_2^{1_p})}(\widetilde{S}) &= -\frac{1}{\sqrt{5}}
\begin{pmatrix}
2 \sin \left(\frac{3\pi}{10}\right) & 2 \sin \left(\frac{\pi}{10}\right) & -\sqrt{2} \\
2 \sin \left(\frac{\pi}{10}\right) & 2 \sin \left(\frac{3\pi}{10}\right) & \sqrt{2} \\
-\sqrt{2} & \sqrt{2} & -1
\end{pmatrix}, \notag \\
\rho_{T^{4}/(\mathbb{Z}_2^{1_t} \times \mathbb{Z}_2^{1_p})}(\widetilde{T}_{I_2}) &=
\begin{pmatrix}
e^{16\pi i/15} & & \\
 & e^{4\pi i/15} & \\
 & & e^{10\pi i/15}
\end{pmatrix},
\label{rho3Gamma15Z3}
\end{align}
which satisfy the following algebraic relations,
\begin{align}
\begin{array}{l}
\ \rho_{T^{4}/(\mathbb{Z}_2^{1_t} \times \mathbb{Z}_2^{1_p})}(\widetilde{S})^2 = I, \\
\ [\rho_{T^{4}/(\mathbb{Z}_2^{1_t} \times \mathbb{Z}_2^{1_p})}(\widetilde{S}) \rho_{T^{4}/\mathbb{Z}_2^{1}}(\widetilde{T}_{I_2})]^3 = I, \\
\ \rho_{T^{4}/(\mathbb{Z}_2^{1_t} \times \mathbb{Z}_2^{1_p})}(\widetilde{T}_{I_2})^5 = e^{4\pi i/3} I.
\end{array}
\label{rho3relGamma17}
\end{align}
Then, they transform non-trivially under $\Gamma_{1,15}$ modular flavor transformation.
Actually, when we define
\begin{align}
s = S, \ t = T_{I_2}^{6}, \ 
c = T_{I_2}^{5},
\label{T'Z2gen}
\end{align}
we can find that they satisfy the following relations,
\begin{align}
s^2=1, \ t^5=(st)^3=c^3=1, \ c x = x c \ (x=s, t).
\label{A5Z3alg}
\end{align}
Thus, they transform non-trivially under $A_5 \times Z_3$ modular flavor transformation.



\end{document}